\documentclass[manuscript]{aastex}

\usepackage{lscape}

\slugcomment{Accepted for publication in the Astronomical Journal}

\shorttitle{A southern CO survey for gas-rich planet-forming disks}
\shortauthors{Hales et al.}

\begin{document}


\title{A CO survey in planet-forming disks: characterizing the gas content in the epoch of planet formation}


\author{A.S.~Hales\altaffilmark{1,2}, I.~De~Gregorio-Monsalvo\altaffilmark{1,3}, B.~Montesinos\altaffilmark{4}, S.~Casassus\altaffilmark{5}, W.F.R.~Dent\altaffilmark{1,3}, C.~Dougados\altaffilmark{6}, C.~Eiroa\altaffilmark{7},  A.M.~Hughes\altaffilmark{8}, G.~Garay\altaffilmark{5}, D.~Mardones\altaffilmark{5}, F.~M\'enard\altaffilmark{6}, Aina~Palau\altaffilmark{9}, S.~P\'erez\altaffilmark{5}, N.~Phillips\altaffilmark{1,3},  J.M.~Torrelles\altaffilmark{10} and D.~Wilner\altaffilmark{11}}

\altaffiltext{1}{Atacama Large Millimeter/Submillimeter Array, Joint ALMA Observatory, Alonso de C\'ordova 3107, Vitacura 763-0355, Santiago - Chile}
\altaffiltext{2}{National Radio Astronomy Observatory, 520 Edgemont Road, Charlottesville, Virginia, 22903-2475, United States}
\altaffiltext{3}{European Southern Observatory, Karl-Schwarzschild-Str. 2, 85748, Garching bei M\"nchen, Germany}
\altaffiltext{4}{Department of Astrophysics, Centre for Astrobiology (CAB, CSIC-INTA), ESAC Campus, P.O. Box 78, 28691 Villanueva de la Ca\~nada, Madrid, Spain} 
\altaffiltext{5}{Departamento de Astronom\'ia, Universidad de Chile, Camino El Observatorio 1515, Las Condes, Santiago, Chile}   

\altaffiltext{6}{UMI-FCA, CNRS/INSU, France (UMI 3386), and Dept. de Astronom\'{\i}a, Universidad de Chile, Santiago, Chile.} 

\altaffiltext{7}{Departamento de  F\'isica Te\'orica, Facultad de Ciencias, Universidad Aut\'onoma de Madrid, Cantoblanco, 28049, Madrid, Spain}
\altaffiltext{8}{Department of Astronomy, University of California, Berkeley, CA 94720, USA}

\altaffiltext{9}{Institut de Ci\'encies de l'Espai (CSIC-IEEC), Campus UAB-Facultat de Ci\'encies, Torre C5-parell 2, 08193 Bellaterra, Catalunya, Spain} 
\altaffiltext{10}{Institut de Ci\`encies de l'Espai (CSIC-IEEC) and Institut de Ci\`encies del Cosmos (UB-IEEC), ~Mart\'{\i} i Franqu\`{e}s 1, 08028 Barcelona, Spain}
\altaffiltext{11}{Harvard-Smithsonian Center for Astrophysics, Cambridge, MA 02138, USA}
\email{ahales@alma.cl}



\begin{abstract}


We carried out a $^{12}$CO(3-2) survey of 52 southern stars with a
wide range of \rm IR excesses ($L_{\rm IR }/L_{*}$) using the single
dish telescopes APEX and ASTE. The main aims were (1) to characterize
the evolution of molecular gas in circumstellar disks using $L_{\rm IR
}/L_{*}$ values as a proxy of disk dust evolution, and (2) to identify
new gas-rich disk systems suitable for detailed study with ALMA.
About 60\% of the sample (31 systems) have $L_{\rm IR }/L_{*} > 0.01$
typical of T-Tauri or Herbig AeBe stars, and the rest (21 systems)
have $L_{\rm IR }/L_{*} < 0.01$ typical of debris disks. We detect
CO(3-2) emission from 20 systems, and 18 (90\%) of these have $L_{\rm
  IR }/L_{*} > 0.01$. However, the spectra of only four of the newly
detected systems appear free of contamination from background or
foreground emission from molecular clouds. These include the
early-type stars HD~104237 (A4/5V, 116 pc) and HD~98922 (A2~III,
507~pc, as determined in this work), where our observations reveal the
presence of CO-rich circumstellar disks for the first time.  Of the
other detected sources, many could harbor gaseous circumstellar disks,
but our data are inconclusive. For these two newly discovered gas-rich
disks, we present radiative transfer models that simultaneously
reproduce their spectral energy distributions and the $^{12}$CO(3-2)
line profiles. For both of these systems, the data are fit well by
geometrically flat disks, placing them in the small class of
non-flaring disks with significant molecular gas reservoirs.


\end{abstract}


\keywords{circumstellar matter -- planetary systems: protoplanetary disks}



\section{Introduction}

Most stars are born surrounded by massive circumstellar disks of dust
and gas \citep[$M_{\rm disk}\sim 10^{-3} -
  10^{-1}M_{\odot}$;][]{williams11}.  As the systems evolve the disks
fade, most of the circumstellar material is accreted either into the
central star, into km-size bodies, or dissipated by other gas dispersal mechanisms 
\citep{2000prpl.conf..401H,2013arXiv1308.1791A}. Models predict that most gas dissipates quickly
after the star is formed \citep[t$<$10 Myr;][]{boss01}, while the
combined effects of collisional grinding of large bodies together with
gravitational resonances can maintain a significant dust grain
population for several Myr \citep[t$<$1Gyr;][]{wyatt08}. 
The fraction of stellar flux re-radiated by the dust (the disk-to-star
bolometric luminosity ratio, $L_{\rm IR }/L_{*}$, also referred as the
infrared excess) is a function of the dust mass present in the
disk. $L_{\rm IR}$/$L_{*}>10^{-1}$ in young T Tauri and Herbig Ae/Be
(HAeBe) stars with massive circumstellar disks, while in more evolved
debris disk systems $L_{\rm IR}$/$L_{*}<10^{-3}$ \citep{zuckerman01}.
Observations indicate that warm (T$>100$~K) dust located in the disk
inner regions (r$<10$~AU) decays with time as t$^{-1}$
\citep{rieke05,meyer07}, as expected from steady-state collisional
evolution models of the dust particles \citep[i.e. the observed dust
  is secondary;][]{wyatt07}, although there are individual systems
with luminosities in excess of steady-state predictions
\citep[e.g. eta Corvi;][]{wyatt05}.

The gas content of protoplanetary disks is more difficult to
probe than dust. 
The zero dipole moment of the most abundant gaseous molecule, H$_{2}$,
hampers its observation. The second most abundant molecule, carbon
monoxide (CO), has a large dipole moment but its observation in
circumstellar disks is very often contaminated by large scale emission,
particularly when observed at low angular resolution
\citep{dent05,vankempen07}. This is unfortunate, as the gas dominates
the disk dynamics in the early protoplanetary stages. The gas-to-dust
ratio affects the thermal and chemical balance
\citep{jonkheid07,2009A&A...501..383W}, as well as the dynamic
coupling of gas and dust \citep{weidenschilling77,takeuchi01}. The gas
temperature determines the scale height of the disk
\citep{dalessio99,2006ApJ...638..314D}, and most importantly, the
mechanisms of gas dispersal will set the timescales for the formation
of gas-giant planets \citep{armitage2011}. Consequently, in the field
of planet formation much effort is currently aimed at constraining the
timescales over which most circumstellar gas dissipates
\citep[e.g.,][]{2010EM&P..106...71C,2011AJ....141..127I,williams11}.

Dynamical instability models predict the formation giant planets on
orbital timescales \citep[t $\ll$ 1~Myr;;][]{2005ApJ...629..535B},
while core accretion models require at least a few Myr to form Jovian
planets \citep{1996Icar..124...62P}. These ages are similar to the age
of most T Tauri and HAeBe systems. Large amounts of molecular and
atomic gas emission are indeed detected towards such objects
\citep{dutrey96,dunkin97,thi01}, but these signatures disappear
towards systems with lower $L_{\rm IR }/L_{*}$ values
\citep{zuckerman95,dent05,moor2011}.  \citet{dent05} carried out a
sub-mm survey for CO emission in a sample of 59 objects with a variety
of $L_{\rm IR }/L_{*}$ values, classified as either T Tauri
and HAeBe stars or more evolved debris disk-like systems. They found
that CO detections are common in objects with $L_{\rm IR}/L_{\rm
  \star}>\,$0.01, but rare in objects with smaller values ($<10\%$ of
these objects). Optical searches for atomic gas over samples with wide
range of $L_{\rm IR }/L_{*}$ also indicate that atomic
emission lines originating close to the stars disappear in stars with
$L_{\rm IR}/L_{*}<\,$0.01 \citep{dunkin97,redfield07},
suggesting a possible transition from gas-rich to gas-depleted disk
near $L_{\rm IR}/L_{*}\sim\,$0.01. However, less than 5$\%$ of
known HAeBe and debris disks like systems have been surveyed for gas
in either CO or optical emission lines \citep{the94,rhee07a},
hampering robust statistical conclusions.

A few $L_{\rm IR}/L_{*}\sim 10^{-3}-10^{-4}$ systems still show
signatures of circumstellar gas, either in the form of CO
emission-line profiles associated with orbiting gas
\citep{zuckerman95,dent05,pascucci2006,moor2011} or by showing sharp
optical absorption features at the core of photospheric lines,
attributable to circumstellar gas around these stars, in the form of
shells, clumps or possibly edge-on disks
\citep{dunkin97,redfield07}. These systems have become of great
interest, as they are thought to be objects transitioning between
their protoplanetary and debris disk phase, close to the end of planet
formation. \cite{hughes08} conducted submillimeter observations of one
of these systems, the A1V star 49~Cet, which revealed a CO gas disk
with a depleted inner region, similar to the structures predicted by
planetary formation and photo-evaporation models \citep{armitage2011,
  owen2012}.  Another interesting 49 Ceti-like system is HD~21997
\citep{moor2011}, for which \citet{kospal2013} recently obtained ALMA
observations. The origin of the gas observed in these systems is still
a matter of debate, as it is not yet clear whether it is primordial
(left from the protoplanetary phase), or secondary \citep[created by
  collisions and/or photodesorption from dust grains,
  e.g.][]{2010ApJ...720..923Z}.

In this work we present a survey for $^{12}$CO (J=3-2) gas over a
sample of 52 southern stars with circumstellar dust excesses selected
from the literature \citep{the94,malfait98,mann98,silv00}, by using
the APEX and ASTE submillimeter telescopes. These two telescopes were
the best southern facilities to provide bona-fide disk detections in
the pre-ALMA era, despite lack of high spatial resolution and
sensitivity. By targeting stars with a range of $L_{\rm IR}$/L$_{\rm
  \star}$ we aim at studying the dependence between L$_{\rm
  IR}$/$L_{*}$ and detectability of CO. No systematic southern
surveys for CO gas in protoplanetary disks have been carried out to
date. \citet{nyman92} used SEST to observe bright southern IRAS
sources, most of which were evolved objects, such as AGB, post-AGB and
carbon stars. Recently \citet{moor2011} conducted a CO search over a
sample of 20 southern debris disks systems, leading to the detection
of the gas-rich debris disk system HD~21997, similar to 49 Cet. A
southern survey is necessary not only to complement results from
northern CO surveys
\citep{zuckerman95,2000MNRAS.312L...1G,dent05,liu11} but also to
provide a sample of gas-rich southern protoplanetary disks that can be
targeted for high-resolution studies with ALMA. Follow-up
high-resolution, high-sensitivity ALMA observations of other gas
tracers can be used to constrain the physico-chemical structure of the
newly discovered disks, as well as to test models predictions in which
certain molecular species will be subject to significant radial
gradients and differentiation \citep[e.g.,][]{walsh2010}.

In Section~\ref{sample} we present our target
selection. Section~\ref{obs} describes our observations and data
reduction.  Our main results are shown in
Section~\ref{results}. Detailed radiative transfer modelling of two of
the CO-rich objects, HD~104237 and HD~98922, is presented in
Section~\ref{rt}. In Section~\ref{disc} we discuss our results and in
Section~\ref{conc} we conclude. Appendix~\ref{cont} presents
individual sources with contaminated CO detections.

\section{The sample}\label{sample}

Our target sample consists of 52 protoplanetary and debris disk
systems selected from the literature as having infrared excesses due
to orbiting dust
\citep{the94,malfait98,mann98,silv00,sylvester00}. The targets were
selected in terms of their photospheric emission line characteristics
(presence or absence of H$_{\alpha}$, i.e. T Tauri/HAeBe versus debris
disk), isolation from molecular clouds and from bright 2MASS sources,
and particularly their $L_{\rm IR}$/$L_{*}$ values.  Most stars
located close to the galactic plane ($|b|<5$) were avoided in order to
minimize cloud contamination in the off- position. We included most
southern HAeBe stars from \citet{malfait98}, and several southern
debris disk systems with $L_{\rm IR}$/$L_{*}>10^{-4}$ from the
compilation of \citet{silv00} and \citet{rhee07a}. The sample
comprised isolated HAeBe and debris disk systems to study the
evolution of the gaseous component of their protoplanetary/debris
disks. The sources of our sample were selected based on different
ratios of the infrared excess flux to the photospheric flux $L_{\rm
  IR}$/$L_{*}$, as a proxy of the evolutionary status of the disks. We
covered a range of $L_{\rm IR}$/$L_{*}$ that varies from several
orders of magnitude, ranging from 10$^{-5}$ to close to unity. We did
not consider objects younger than 10$^6$ years nor embedded objects,
to avoid contamination from protostellar envelopes or from molecular
cloud material. Thus, with these proposed `naked' disks, we expect to
reduce the possibility of CO contamination by their parental
clouds. Table 1 lists our targets, their spectral types and distances,
as well as their $L_{\rm IR}$/$L_{*}$ values. \\

\begin{center}
\begin{figure}
\includegraphics[angle=0,scale=0.55]{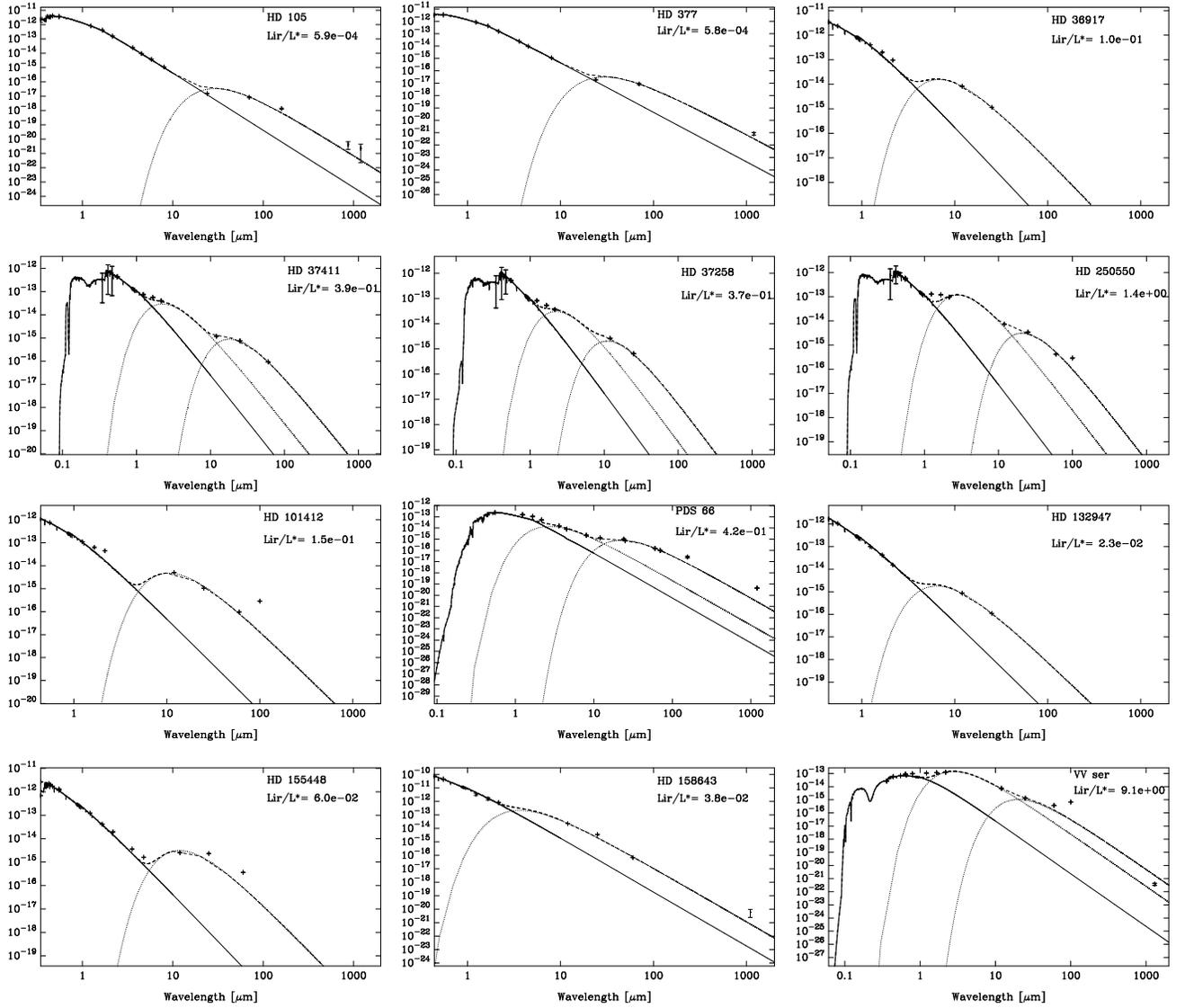}
\caption{Example of our blackbody fit to the available photometry in
  order to derive $L_{\rm IR}$/$L_{*}$ values for stars for which
  infrared excess were not found in the literature. Units in the
  vertical axis are in erg/cm$^{-2}$/s.  }\label{figsedfits}
\end{figure}
\end{center}

The $L_{\rm IR}$/$L_{*}$ values listed in Table~1 were taken from the
literature \citep{acke04,meeus12}, when available, or by fitting
black-bodies to published optical, infrared and sub-mm data from the
literature. The observed SEDs are assumed to be the sum of the model
stellar atmosphere SED (SED$_{\star}$) and a cooler black body of a
given temperature (SED$_{\rm \small disk }$),
i.e. $SED_{tot}=SED_{\star}+SED_{\rm \small disk}$. For each stars in
Table~1, we searched for the model that minimised the chi-squared
($\chi^{2}$) difference between the modelled SED$_{tot}$ and the
observed data-points.  The search was performed using the
variable-metric routine Migrad of the Minuit package from
CERN\footnote{\tt
  http://seal.web.cern.ch/seal/work-packages/mathlibs/minuit/home.html}.
The photospheric luminosities (SED$_{\star}$) were computed from stellar model
atmospheres (assuming spectral types listed in SIMBAD), which were
reddened to match the optical data. A value of $R_{\rm V}=3.1$ was
assumed. It is well known that the blue and ultraviolet regions of the
spectrum are the ones most affected by extinction, therefore, when
available, ultraviolet spectra obtained with the {\it{IUE}}
\citep[International Ultraviolet Explorer;][]{1978Natur.275..372B}
observatory were added to the SED; they were extremely useful in many
cases to estimate $E(B-V)$. L$_{\rm IR}$/$L_{*}$ was then obtained by
fitting black-body laws to the infrared excess. In some cases two
black-bodies were required in order to fit both the near-IR and far-IR
excesses. We estimate our method to be accurate by factors of 2-3 by
comparing our $L_{\rm IR}$/$L_{*}$ results with stars with values
already published in the literature \citep{sylvester00,meeus12}.
Example results from our SED fitting method are shown in
Figure~\ref{figsedfits}.

In our sample 59.6~$\%$ of the sources have L$_{\rm IR}$/$L_{*}$
values typical of HAeBe and T Tauri stars ($>0.01$, for which
\citet{dent05} find a 48\% CO detection rate). The other 40.4~$\%$ of
the objects have $L_{\rm IR}$/$L_{*}<0.01$ and have been selected in
order to characterize the transition zone where gas-rich to
gas-depleted disks is thought to occur \citep[L$_{\rm
    IR}$/$L_{*}\sim0.01$;][]{dent05}. The Herbig Ae star HD~100546,
known to harbor a CO-rich protoplanetary disk, was also included in
the sample as a control target \citep{2010A&A...519A.110P}.



\section{Observations and data reduction}\label{obs}

Here we describe the main $^{12}$CO(3-2) single dish survey, and
complementary far-infrared and submillimeter line and continuum
observations of a few of the detected sources.

\subsection{ASTE and APEX $^{12}$CO(3-2) observations}

Observations were carried out using the Atacama Submillimeter
Telescope Experiment \citep[ASTE;][]{kohno05} and the Atacama
Pathfinder Experiment\footnote{Based on observations with the Atacama
  Pathfinder EXperiment (APEX) telescope. APEX is a collaboration
  between the Max Planck Institute for Radio Astronomy, the European
  Southern Observatory, and the Onsala Space Observatory.}
\citep[APEX;][]{gusten06}, both located at 5000~m above sea level in
the Atacama Desert. Spectra towards each star were acquired using the
standard on-off observing pattern to remove atmospheric continuum
emission, using typical beam-switching offsets of 180 arcseconds in
azimuth. In cases where spurious absorption lines were seen in the
spectra due to ambient emission in the off- position, a second
observation using an off- position of $-180$ arcseconds in azimuth was
obtained. Spectral line observations of the J=3-2 transition of
$^{12}$CO (rest frequency $\nu=345.795991$~GHz) were made using the
2-SB 350~GHz receiver on ASTE and the SHeFI 345~GHz receiver on
APEX. The velocity resolution on ASTE and APEX was of 0.5~km~s$^{-1}$
and 0.4~km~s$^{-1}$,respectively. The total bandwidths covering the
$^{12}$CO line were 1~GHz and 0.5~GHz for APEX and ASTE respectively.
The beam widths of APEX and ASTE at this frequency were 18 and 21
arcseconds, respectively. Typical disk sizes for protoplanetary disks
are a few hundred AU, and will therefore be unresolved in our
observations. The corresponding main beam efficiencies ($\eta_{\rm
  mb}$) listed in the ASTE\footnote{\tt
  http://www.nro.nao.ac.jp/$\sim$aste/cfp2011/note.html} and APEX\footnote{\tt
  http://www.apex-telescope.org/telescope/efficiency/} instrument webpages are
0.6 and
0.73.

In order to reach a targeted root-mean-square (RMS) noise level of
$0.02\,-\,0.04\,$~K \citep[comparable to the northern survey of][]{dent05},
typical integration times of 1 and 2 hours were required for APEX and
ASTE, respectively. The survey was completed in different observing
blocks during June 2008, July 2009, December 2009, and July 2010 (APEX
Project Codes C-081.F-0010B, C-083.F-0187A, and C-084.F-0478A). The
precipitable water vapour was typically between 1mm and 2mm, with
exceptionally good conditions ($0.3\,-\,0.9\,$mm) encountered during the
July 2010 run. Close to 80 hours of telescope time were required for
completing the survey.

The ASTE spectra were converted to FITS format using the Newstar
package\footnote{\tt
  http://www.nro.nao.ac.jp/~nro45mrt/obs/newstar/}. After conversion
to FITS, both the ASTE and APEX spectra were processed using the
GILDAS/CLASS reduction software \citep{guilloteau00}\footnote{\tt
  http://www.iram.fr/IRAMFR/GILDAS/doc/html/class-html/class.html}. The
raw data individual scans were visually inspected for identifying
spikes, outliers, and baselines instabilities.  Scans with problems
were excluded from the data reduction.  A baseline substraction from
each individual scan was performed excluding channels that showed
spectral emission line. A polymonial of degree three or less was used
for the baseline fitting.  In order to produce the final spectra with
the highest signal -to-noise ratio, scans were averaged.  The rms
noise was computed in the averaged spectra excluding the channels that
contained signal from the object of interest.

The 2008 ASTE data were affected by an instrumental 1~MHz
sinusoidal ripple of variable amplitude and phase. This artifact was
removed from each individual 10~s integration by fitting the first
peak of the spectrum's Fourier components and subtracting it
\citep[][for details]{hughes2010}.

\subsection{LABOCA Archive observations}

An $870\,\mu{\rm m}$ continuum map of the HD~104237 system obtained
with the LABOCA bolometer array camera on APEX \citep{2009A&A...497..945S}
was retrieved from the ESO archive\footnote{\tt
  http://archive.eso.org} (PI R. Liseau, project code 082.F-9304A). The
observations were performed on 28 December 2008 in stable conditions
with line-of-sight $\tau=0.38$, using a standard compact raster of
spiral scan pattern which yields excellent spatial
sampling. Instrumental sensitivity, as determined by flux calibrator
measurements, was reduced to approximately 64\% of nominal this night
due to a bias problem. The data were scaled to compensate for this in
the reduction, but the noise is accordingly increased. 

Data were reduced using a custom pipeline in the BoA
package7 \footnote{http://www.apex-telescope.org/bolometer/laboca/boa/}.
The pre-processing steps consisted of flagging dead or cross-talking
channels, frames with too high telescope accelerations and with
unsuitable mapping speed, as well as temperature drift correction
using two blind bolometers.  Data reduction process included
flat-fielding, opacity correction, calibration, correlated noise
removal (atmospherics fluctuations seen by the whole array, as well as
electronic noise originated in groups of detector channels), and
de-spiking.  Every scan was visually inspected to identify and discard
corrupted data. The final map shows an angular resolution of ~19" and
an RMS noise of 18 mJy/beam (calculated in the central part of the
map).


\subsection{{\it Herschel}/PACS Archive Observations}

We have analysed unpublished Herschel observations of HD~98922 using
the PACS instrument in both photometric and spectroscopic modes that
exist in the {\it Herschel} Science Archive (Pointed Range
Spectroscopy observing mode).  We used the Level 2 archived standard
products produced by the Herschel Standard Product Generation (SPG)
software, version 10.3.0.  PACS photometry was extracted by PSF
fitting to obtain 70 and 160~$\mu$m fluxes (3815.2 877.5~Jy
respectively). The RMS in the 70 and 160~$\mu$m maps are 0.2 and 0.4
mJy/pixel respectively. The uncertainty in the derived fluxes will be
dominated by the 5\% absolute calibration accuracy of PACS
\citep{balog2013}, so we quote 5\% photometric errors for the PACS
final fluxes in Table~\ref{tablefluxes2}.


\section{Results}\label{results}

Figure~\ref{figco1} shows the spectra for four detections that appear
to be free from foreground/background contamination in the off-
spectra. Association of the CO emission to the targeted stars may be
determined by comparing the velocity of the detected emission to the
velocity of the star determined from optical spectra (the known
$V_{\rm LSR}$ velocity of the stars are plotted on top of the spectra,
when available).\\

Integrated line intensities on the detections, or 1-$\sigma$ upper
limits, are presented in Table 1 in antenna temperature units. Upper
limits were computed as $T_{rms}\Delta v \sqrt{N}$, where $T_{rms}$ is
the noise measured in the spectra, $\Delta v$ is the velocity
resolution of each channel, and $N$ is the number of channels over a
10~km s$^{-1}$ velocity interval. We report an integrated CO intensity
of 2.5~$\pm0.3$~K~km~s$^{-1}$ ($T_A^\star$) towards our control target
HD~100546, compared to the previously measured
2.9~$\pm0.4$~K~km~s$^{-1}$, which are consistent within the
uncertainties.  \citep[also in $T_A^\star$
  ;][]{2010A&A...519A.110P}. \\

A single-dish non-detection of CO implies that neither a CO disk nor
CO ambient emission is detected towards the targeted star (fainter CO
disk could be present below the sensitivity limit).

 However, a CO detection does not necessarily mean that a
gaseous disk is present. \citet{vankempen07} already noted the
difficulty of disentangling ambient CO emission versus protoplanetary
disk signatures when looking for circumstellar CO using a single-dish
telescopes towards pre-MS stars in Lupus.  While most debris disk
systems are isolated from molecular clouds, younger stars are usually
still associated to their parent clouds (see
Figure~\ref{figmilkyway}), increasing the probability of contamination
from ambient CO in the spectra (either in the off-, the on- position,
or both), and need interferometric observations to filter out the
extended cloud emission. \\

We surveyed 52 stars for $^{12}$CO(3-2) emission and obtained
detections towards 20 objects. 18 (90$\%$) of the detected sources
have $L_{\rm IR}$/$L_{*}>$0.01.  This trend is in agreement with the
findings of the survey from \citet{dent05} in which $\sim50\%$ of the
systems with $L_{\rm IR}$/$L_{*}>\,0.01$ have significant CO
detections. Our statistics are, however, biased due to cloud
contamination toward many of the sources. The description of the
individual sources and results with detections with
background/foreground contamination is given in Appendix~\ref{cont}.


\subsection{$^{12}$CO Disk detections}

\subsubsection{HD~104237} 


HD~104237 (DX Cha) is the only early-type star of the young $\epsilon$
Cha cluster \citep[$3-5$~Myr;][]{feigelson03}. It is a Herbig Ae star
\citep[A4Ve-A8Ve;][]{grady04,lyo2008} with infrared excess
\citep{walker88,hu89}. At least five stars are found within 15
arcseconds of HD~104237, labeled with letters from "B" to "E"
\citep{feigelson03}. They are lower-mass stars of which at least two
have IR excesses and emission lines, associated to T Tauri type stars
\citep{grady04}. Optical spectroscopy has revealed that HD~104237 is
actually a close binary system, where the secondary is a K3 star
(HD~104237-2) at $\sim 0.2$~AU \citep{bohm04}.  Model fits to its SED
suggest that HD~104237 is surrounded by a dust disk with an irradiated
inner wall that shadows the outer disk \citep{meeus01}.  Very Large
Telescope (VLT) AMBER observations resolve the inner radius of the
ring at R$\sim0.45$~AU \citep{tatulli07,garcia2013}. The binary
separation $a = 0.22$~AU implies the disk is circumbinary
\citep{bohm04}.

HD~104237 belongs to a small group SED-predicted flat disk sources
that exhibit moderate [O I] emission at optical wavelengths
\citep{acke05}.  Recent modeling of the optical to mid-infrared SED
by \citet{fang2013} supports \citet{meeus01} SED interpretation of
HD~104237's being surrounded by a self-shadowed (i.e. non-flared) dust
disk \citep[therefore being a Group II source according to the
  classification of][]{meeus01}.

There is evidence of circumstellar gas around HD~104237 detected in
the Br$\gamma$ line \citep{kraus08}, and marginally detected in the
optical [O I] 6300 {\AA} line \citep[the detection is contaminated by
  photospheric absorption;][]{acke05}. While an [O I] emission could
arise from warm surface layers of an orbiting gas disk, the hydrogen
Br$\gamma$ emission appears to come from a gaseous inner ($<1$ AU)
disk \citep{kraus08}. \citet{garcia2013} suggest, from recent VLT
data, that Br$\gamma$ is actually very compact and mostly arises from
the close environment of each component of the binary, probably
related to accretion flows. However, these findings are derived from
interferometric observations that filter out any component extended
over scales larger than a few AUs. No H$_2$ emission was detected in
the pure rotational line in the mid-IR
\citep{carmona08}. \citet{meeus12} detected strong [O I] 63.2~$\mu$m
emission towards the HD~104237 using PACS on the {{\it Herschel Space
    Telescope}}, indicating that there are significant amounts of
circumstellar gas around this system. A large-scale jet elongated in
the southeast-northwest directions and seen in Ly$\alpha$ originates
from this source \citep{grady04}.

A double-peaked CO line profile is detected in the ASTE spectrum
(Figure~\ref{figco1}), centered at 5~km~s$^{-1}$ and of roughly
8-9~km~s$^{-1}$ width. 
HD~104237 is located in a region between the Cha II and Cha I clouds,
at the edge of Cha II. The NANTEN $^{12}$CO(1-0) map \citep{mizuno01}
shows ambient emission close to the HD~104237 system in the
2-6~km~s$^{-1}$ range. The stellar velocity plotted on top of the
spectrum was obtained from \cite{2007yCat..90320844G}. The CO emission
is centered at the star's position (confirmed by spectra with
different off-positions) , indicating that the disk contribution
dominates the ambient cloud contribution.

\subsubsection{HD~98922} 

Together with HD~104237, HD~98922 is also member of the small groups
flat-disk sources with [O I] emission at optical wavelengths
\citep{acke05}.  No H$_2$ emission was detected in the pure rotational
line in the mid-IR \citep{zaidi08}.  Similarly to HD~104237, the 63.2
$\mu$m [O~I] line is also detected \citep{fedele2013}.\\

A single-peaked profile is detected in CO(3-2) towards HD~98922
(Figure~\ref{figco1}), centered at -7~km~s$^{-1}$. The line velocity
is offset by 15~km~s$^{-1}$ from the radial velocity listed in
\cite{acke05}. This offset is, however, comparable to their
centroiding error in the stellar velocity determination (between 8 and
$-15$~km~s$^{-1}$).  HD~98922 is located at 7 degrees above the
Galactic Plane ($(l,b)=\,$(289.78, +07.21)) in the direction of the
Third Quadrant \citep{dame01}, but there are no prominent known
molecular clouds on the line-of-sight.  The two closest clouds derived
from extinction maps are located at $(l,b)=$(290.70, 5.83) and
(290.93, 7.50) respectively \citep{dobashi05}.  Since there are no
known molecular clouds in the line-of-sight, nor evidence of
contamination in the reference spectrum, we associate the CO emission
to the stellar position (within the ASTE primary beam).


\subsubsection{HD~155448} 

Classified as a B9-type star, there have been two main interpretations
in the literature concerning the nature of HD~155448: it has been
proposed to be a HAeBe star with a circumstellar disk
\citep{malfait98,schutz04}, or a post-AGB star surrounded by an
ejected shell \citep{luna08,ortiz05}.

\citet{schutz2011} seems to have resolved this dichotomy by analysing
the system in detail using resolved photometry and spectroscopy. They
propose this is actually a quintuple system and rule out a possible
post-AGB nature: 4 out of 5 components lie close to the ZAMS in the HR
diagram. According to \citet{schutz2011}, the C component presents
spectral signatures corresponding to a star with a circumstellar disk.

A single-peaked profile is detected towards HD~155448, centered at
24~km~s$^{-1}$ and of roughly 6~km~s$^{-1}$ width
(Figure~\ref{figco1}). \citet{luna08} detected optical absorption
features of H$_\alpha$ and Carbon at 36~km~s$^{-1}$ and 26 km s$^{-1}$
respectively, which they associated to an expanding circumstellar
envelope. Even though the expanding post-AGB envelope scenario has
been ruled out by \citet{schutz2011}, it is interesting to note that
the $^{12}$CO(3-2) emission of HD~155448 coincides in velocity with
the optical atomic carbon emission detected in \citet{luna08}.

\subsubsection{RX~J1842.9-353} 

A double peaked profile characteristic of rotating disks is detected
towards RX~J1842.9-353 in the ASTE spectrum (Figure~\ref{figco1}). The
ASTE detection was analyzed alongside complementary SMA data presented
in \citet{hughes2010}.  In the SMA data the dust disk was detected and
marginally spatially resolved, while the CO observations resulted in an
interferometric upper limit.

%
%
%



\subsection{LABOCA $870\,\mu{\rm m}$ continuum map of HD 104237}
  
\begin{center}
\begin{figure}
\includegraphics[angle=0,scale=1.2]{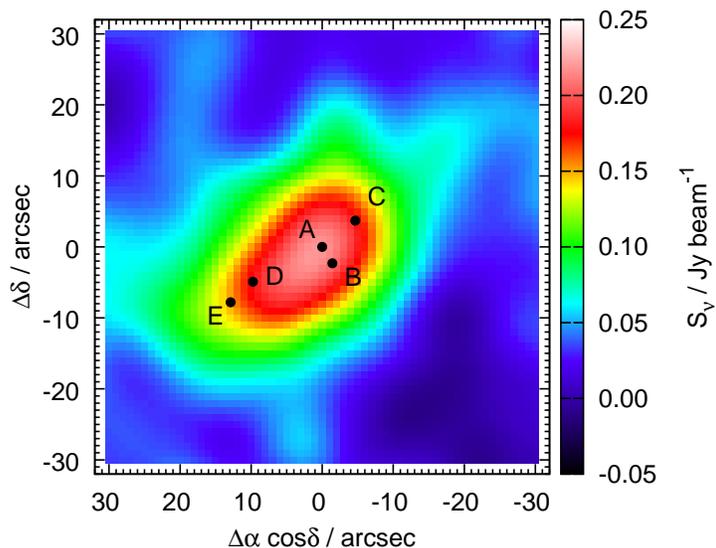}
\caption
{LABOCA map of HD~104237 smoothed with half-beam Gaussian (${\rm
    FWHM}=9.3''$), with stellar component positions plotted (component
  A corresponds to HD~104237).  }
\label{figlaboca1}
\end{figure}
\end{center}

\begin{center}
\begin{figure}
\includegraphics[angle=0,scale=1.2]{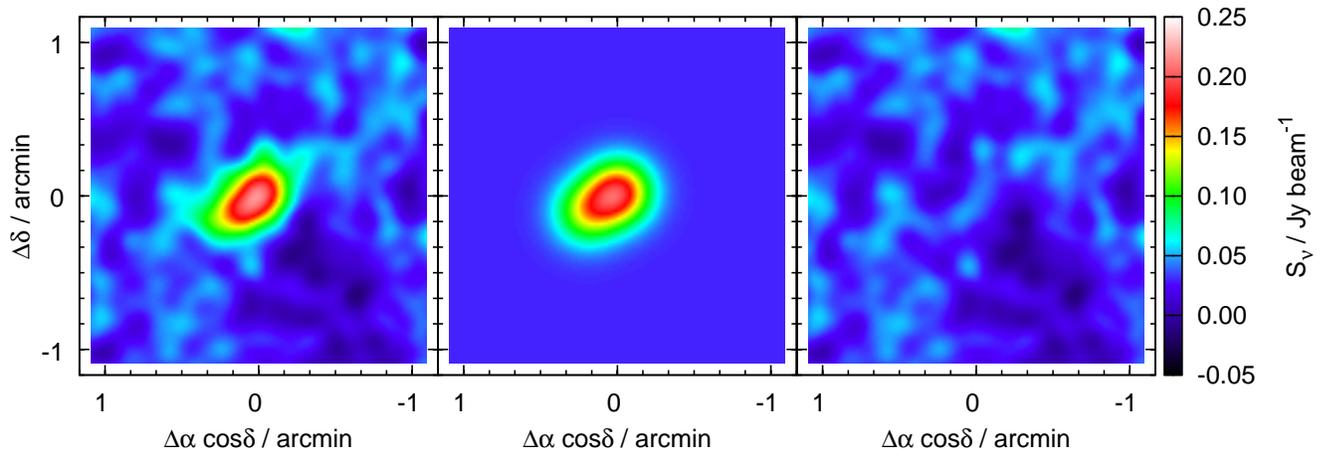}
\caption
{{\it{Left:}} LABOCA map; center: fitted model of 2 effective beam-sized Gaussians (${\rm FWHM}=18.6$) with separation of A and E components; {\it{Right:}} maps with model subtracted.   }
\label{figlaboca2}
\end{figure}
\end{center}

The LABOCA $870\,\mu{\rm m}$ continuum map of HD 104237
(Figure~\ref{figlaboca1}) shows a clear detection of the primary disk,
but also significant extension to the South-East, consistent with
emission from a disk around the D or E components in the system. Based
on the mid-IR detection of a disk around the E component by
\citet{grady04}, we have extracted fluxes using a model of two point
sources separated by the $14.9''$ A-E separation vector
(Figure~\ref{figlaboca2}). This yielded fluxes of $154\pm24\,{\rm
  mJy}$ and $91\pm24\,{\rm mJy}$ for the A and E components
respectively (uncertainties allow for S/N, calibration uncertainty and
anti-correlated error of the two fluxes), and a position offset error
of under $2''$. Due to the system configuration, any flux from the D
component will mostly be incorporated in the E component flux, and any
emission from the B and C components will be included in the A
component flux.


%
%
%
%
%

%

\section{Radiative transfer modeling of HD~104237 and HD~98922}\label{rt}

In this section we use the 3D radiative transfer Monte Carlo code
MCFOST \citep{pinte06} to model simultaneously the SEDs and
$^{12}$CO(3-2) spectra of two of the four sources exhibiting emission
from a disk and not contaminated by known molecular clouds. We model
only two sources, HD~104237 and HD~98922. RX~J1842.9-353 is already
modeled by \citet{hughes2010}, and given the complexitiy of the
HD~155448 system \citep{schutz2011}, we do not carry out radiative
transfer of this system but do include it in our final detection
statistics.

The models presented in this section assume that the dust and gas are
well coupled, with a dust-to-gas ratio of 10$^{-2}$ and a 10$^{-4}$ CO
abundance with respect to H$_2$. The canonical average dust-to-gas
ratio found in the literature for the ISM is in the order of
10$^{-2}$, we assume that young protoplanetary disks should then have
similar ratios, although decreasing as the gas dissipates and dust
grains grow \citep{Birnstiel2010}. The relative abundances of CO and
H$_2$ are constant in the diffuse ISM and equal to 10$^{-4}$
\citep{Federman1980,lee1996}. These standard values for the
dust-to-gas and CO/H$_2$ ratios therefore constitute a first
reasonable approach to the true physical properties of the disks.

MCFOST assumes that the gas-dust thermal exchange is perfect and that
gas and dust are in local thermodynamic equilibrium (LTE) throughout
the disk. The LTE approximation is justified for low-J CO lines, as
has been demonstrated several times
\citep{Pavlyuchenkov2007,panic2009}.  Freeze out is a natural outcome
of a cold media containing dust and gas. It is well known from
laboratory and from astronomical data that CO freezes out at
$\sim$~20~K \citep{qi2011,degregoriomonsalvo}. Models accounting for
the freeze-out of CO into dust grains at temperature of $\sim$~20~K
can also be incorporated in MCFOST.

\subsection{HD104237 SED and CO disk modeling with MCFOST}\label{rthd104237}

\begin{center}
\begin{figure}
\includegraphics[angle=0,scale=0.6]{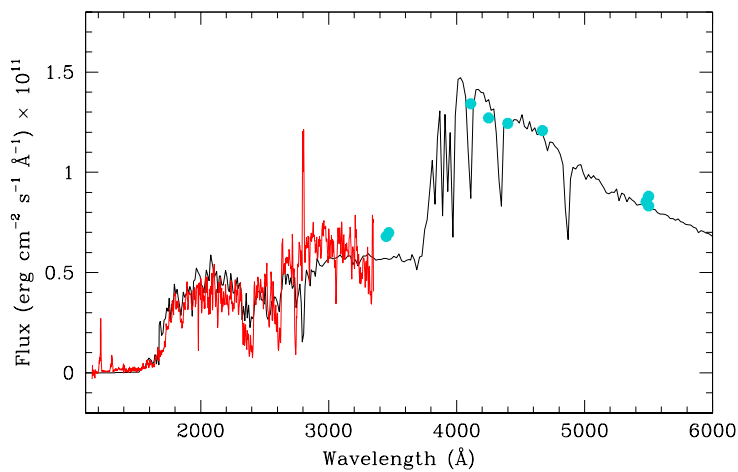}
\caption
{ HD~104237 spectral type fit to the {\it{IUE}} (red line) and optical
  data (blue dots). The synthetic spectra (black line) corresponds to
  the sum of two Kurucz models (T$_1=$8000, $\log(g_1)=4.0$ and
  T$_2=4750$ , $\log(g_2)=$4.5) ;Kurucz 1993), where the second model
  was scaled such as the integral of the first one divided by the
  integral of the second one is 10. The combined model has been
  computed adding up these two files. The combined model has been
  normalized to the flux at V with no reddening and with
  E(B-V)=0.05. The best fit does not seem to require any reddening.  }
\label{figifu}
\end{figure}
\end{center}

Due to its complexity (circumstellar activity, multiplicity), the
basic stellar parameters of HD~104237 have been been loosely
constrained. There is a large discrepancy in temperature, luminosity
of the primary (HD~104237 or HD~104237A hereafter) and the secondary
(HD~104237A-2) star \citep[e.g. for the primary T$\sim7300-9550$~K,
  and L$\sim20-60$~L$_\odot$;][]{fumel2012}. Table ~\ref{tablefluxes1}
summarizes the available photometric data used for the SED modelling
of this system. For our SED fitting we adopted the effective
temperatures and luminosities from \citet{tatulli07}, that is
T$_{eff,1}=8000$~K, L${_1}=30$~L$_{\odot}$, for the primary and
T$_{eff,2}=4750$~K, L${_2}=3$~L$_{\odot}$ for the secondary star. This
combination of stellar parameters is able to fit the available optical
and {\it{IUE}} data \citep{1978Natur.275..372B} with no need of any
reddeding (Figure~\ref{figifu}), and were therefore adopted as the
energy source for the radiative transfer modeling. Other parameters
such as stellar masses and stellar radii are taken from
\citet{garcia2013}.  The fundamental stellar parameters assumed for
modeling the energy input of the system are presented in
Table~\ref{tablemodel1}.\\


The surface density distribution of the disk that was adopted is the
tapered-edge model \citep{andrews2009,williams11},

\begin{eqnarray}
 \Sigma=\Sigma_C \left( \frac{R}{R_C} \right)^{-\gamma} \exp \left[- \left( \frac{R}{R_C}\right)^{2-\gamma} \right] 
\end{eqnarray}

, where $\Sigma_C$ is the surface density at the characteristic radius
R$_{C}$, and $\gamma$ is a variable physically related to the disk's
viscosity \citep{2008ApJ...678.1119H}. The latter controls the radial
variation of the density distribution by providing a power law
distribution in the inner disk with an exponential taper in outer
regions. The inner rim value of 0.45~AU was adopted from
\citet{tatulli07}. The flaring parameters were also varied, assuming
that at at given radius the disk column density is distributed
vertically like a Gaussian with a scale-height

\begin{eqnarray}
 H=H_cR_c\left(\frac{R}{R_C}\right)^{\psi}
\end{eqnarray}

where H$_C$R$_C$ is the scale height at the characteristic radius
R$_{C}$, and $\psi$ describes the dust disk flaring.  In summary, the
total disk mass (M$_D$, obtained by integrating the entire density
distribution), R$_{C}$, $\gamma$, $\psi$ and $h_C$ (the scale height
at the characteristic radius) were variables of the model fitting
routine.

%
The dust grains consist of astronomical silicates and are distributed
in size following a power law, $n(a)\propto a^{-3.6}$ between
$a_{min}=0.03\,\mu$m and $a_{max}=7000\,\mu$m
\citep{draine84}. Without constraints on the disk sizes (i.e. direct
imaging), the fitting process is a highly degenerate problem: changes
in the disk outer radius and the surface density distribution can be
made to artificially math the SED and CO data. A range of $\gamma$
values from 0 to 2 were explored in order to fit the SED data, until a
good match was found. A range of disk outer radius from 60 to 120~AU
was explored. The disk'a total mass was mostly constrained by the long
wavelength data points. However, given the uncertainties and
assumptions in the models, we did not attempt to optimise these fits
any further and therefore no formal errors are quoted in the variable
terms. We stress the fact that this model is simply a rough model that
matches the data and is by no means a best-fit model in formal terms.
the data and is by no means a best-fit model in formal terms.


%
%

The SED can be fit with a 90~AU disk of total dust mass
$4\times10^{-4}\,$M$_{\odot}$. The extension of the fitted disk is
comparable with the 70~AU disk radius estimated by
\citet{grady04}. The total mass we derive is in agreement with that of
\citet{henning1994}, who estimated the mass of the system based on the
1.3~mm data. \citet{fang2013} estimates a smaller disk masses in their
models, but with a smaller wavelength range ($<$20~$\mu$m) their SED
fits were not sensitive to larger dust grains. No flaring is required
to match the SED ($\psi=0$), in good agreement with previous SED
studies of the disk's vertical structure which indicate the disk is
not flared \citep{meeus01,fang2013}. We use this fiducial model to
attempt to also reproduce the CO observations.

The degeneracies when mining pure SED modeling lacking resolved images
are well known, but nonetheless the SED information and the kinematic
information from the gas lines can be used to further constrain the
disk geometry \citep{dent05,panic2009,hughes2010}. MCFOST uses the
radiation and temperature fields (assuming T$_{dust}=$T$_{gas}$) to
compute the level populations of CO (assuming LTE), and to produce line
emission surface brightness profiles. We have used a gas-to-dust ratio
of 100:1 and abundance of CO of 10$^{-4}$ respective to H$_2$,
constant throughout the disk.\\

\citet{grady04} estimates a disk inclination angle of
$18_{-11}^{+14}$~degrees. This value was used as initial estimate but
was then allowed to vary during the different runs. The velocity field
is that of Keplerian rotation with a central object of mass equal to
that of the combined binary system (M$_1$+M$_2$=3.6~M$_{\odot}$). The
systemic velocity ($\rm{V}_{\rm{lsr}}$) and the inclination of the
disk were varied until a good match was found.

\begin{center}
\begin{figure}
\includegraphics[angle=-90,scale=0.5]{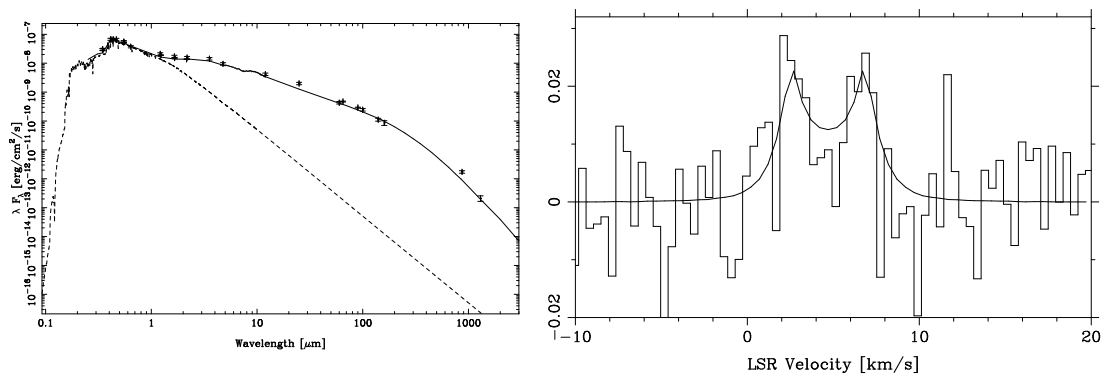}
\caption
{ {\it{Left:}} Observed SED of HD~104237 (black points with error
  bars) plotted against the MCFOST disk model that fits the optical to
  sub-mm data (solid line) . The full photometric data is available in
  the online version of the article. {\it{Right:}} MCFOST CO model of
  HD~104237 (solid line) fits well the ASTE sub- mm data. Intensity
  units are in antenna temperature (T$^{*}_{A}$).}
\label{hd104237-lime}
\end{figure}
\end{center}

We find that the standard model over predicts the flux densities by a
factor of 2.5. As discussed in \citet{hughes2010}, there are several
ways of reducing the predicted CO intensities without changing the
dust distribution.  Different CO/H$_2$ ratios, dust-to-gas ratio, or
different gas temperatures would significantly change the observed
intensities. By incorporating the freeze-out of CO molecules into dust
grains at temperatures below 20~K (that is the abundance of CO drops
to zero where $T<20$~K in the disk), the model images fit the data
without the need of changing neither the global abundance of CO/H$_2$,
nor the disk total mass. The parameters that reproduce the CO data
(Figure~\ref{hd104237-lime}) are listed in Table~\ref{tablemodel1}.

\subsection{HD~98922 SED and CO disk modeling with MCFOST}

Contrarily to HD~104237, very little information is available on
HD~98922. Together with HD~104237, it is one of the
few close-by, southern, flat disks with signs of a rich gas chemistry.

The determination of the parameters for HD~98922 turned out to be a
particularly difficult issue. The star is bright ($V\!\sim\!6.8$) and
was classified as a B9~Ve (Houk, 1978). Taking the typical absolute
magnitude for a B9 V star $M_V=+0.2$, that would imply a distance of
about 210~pc, whereas the Hipparcos parallax $\pi\!=0.98\pm0.39$ mas (van
Leeuwen, 2007) puts the star much further, leading to a contradiction.

The spectral type B9~Ve seems to have been used in many subsequent
works without revision. Following the hint given by the bad fit of the
ultraviolet IUE~spectrum using models with $T_{\rm eff}\simeq
10500$~K, typical of a B9AV, it turns out that a temperature around
9000 K gives more sensible results. Garcia-Lopez at al. (2006)
reported a fairly high accretion rate $\log \dot{M_{\rm acc}}$
($M_\odot$/yr)=$-5.76$, so one has to be careful interpreting the UV
spectrum.

An UVES/VLT spectrum (Mario van den Ancker, private communication) was
used to estimate the stellar gravity, using the width of the Balmer
lines, and the metallicity. The conclusion is that a model with
$T_{\rm eff}$=9000~K, $\log g_*$=3.0 and [Fe/H]=$-0.5$, fits
reasonably well the spectrum --and also the SED-- of the star. This
set of parameters would correspond to a spectral type around
A2~III. Typical errors would be $\pm 250$~K in $T_{\rm eff}$, $\pm
0.1$ dex in $\log g_*$ and $\pm 0.3$ dex in [Fe/H].  Concerning the
metallicity, which is a delicate issue, Folsom et al. (2012) point out
the fact that HD~98922 shows suspiciously weak metal lines for their
literature temperature, hinting at a potential presence of $\lambda$
Boo peculiarities. Although that analysis is out of the scope of this
work, a fit of the region 6140--6180 \AA, which contains three O {\sc
  i} (6156.0, 6156.8 and 6158.2 \AA) and two Fe {\sc ii} lines shows
that whereas the Fe {\sc ii} features are well reproduced with an
abundance [Fe/H]=$-0.5$, or even lower, the oxygen abundance --scaled
to the iron abundance-- seems to be much larger.

\begin{center}
\begin{figure}
\includegraphics[angle=0,scale=0.7]{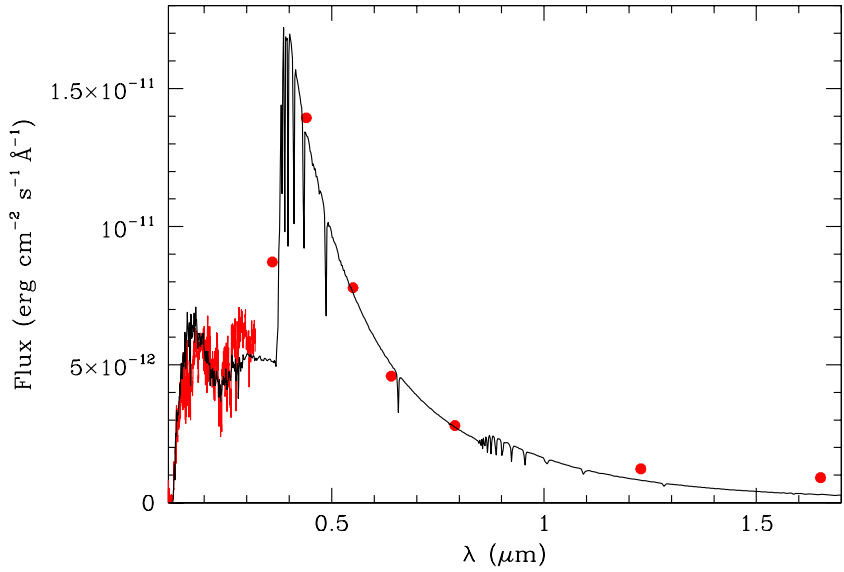}
 \caption
{
 The observed spectral energy distribution of HD 98922 (red
 line and dots) and the fit of a low-resolution Kurucz model (black)
 computed with $T_{\rm eff}\!=\!9000$ K, $\log g_*\!=\!3.0$,
 [Fe/H]=$-0.5$ reddened with $E(B-V)=0.07$ ($R_V\!=\!3.1$) normalized
 to the flux at $I$. The ultraviolet spectrum is a composite of the
 IUE spectra SWP 18553, SWP50065, SWP50083, LWR14624, LWP30007,
 LWP30026 that were retrieved from the archive at the CAB Science Data
 Centre (\tt{http://sdc.cab.inta-csic.es/ines/ }). 
}
\label{HD98922_SED}

\end{figure}
\end{center}

\begin{center}
\begin{figure}
\includegraphics[angle=0,scale=0.7]{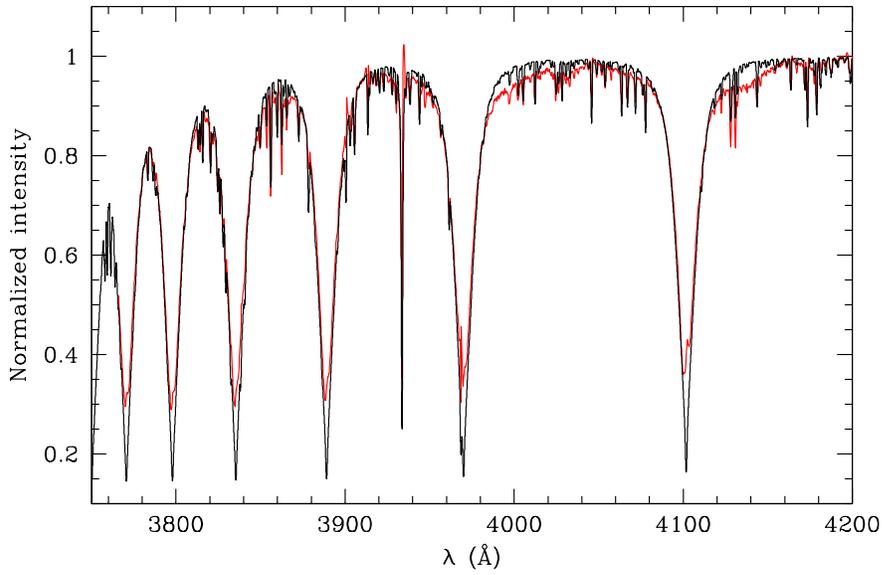}
\caption 
{The region 3750--4200 \AA{} of the spectrum of HD 98922 obtained with
  UVES/VLT (red) and the synthetic spectrum (black) computed with the
  Kurucz ATLAS9 and SYNTHE codes for $T_{\rm eff}\!=\!9000$ K, $\log
  g_*\!=\!3.0$, [Fe/H]=$-0.5$ (see text for details).}
\label{HD98922_SP}
\end{figure}
\end{center}

Figures~\ref{HD98922_SED} and \ref{HD98922_SP} show the fits to the
ultraviolet and optical observations and the spectral fit to the
Balmer lines H$\delta$, H$\epsilon$, H$_8$ -- H$_{11}$. Low resolution
synthetic spectra from the grid of Castelli-Kurucz and high-resolution
synthetic spectra computed using the ATLAS9 and SYNTHE codes by Kurucz
(1993) were used.

Using the results from the spectral and SED fits we can obtain an
estimate of the distance. The process is explained in detail in
\citet{montesinos2009} and makes use of the position of the star in
the HR diagram $\log g_*$ -- $\log T_{\rm eff}$, which can be
translated to a point in the HR diagram $\log L_*/L_\odot$ -- $\log
T_{\rm eff}$ to obtain an estimate of the luminosity. Once
$L_*/L_\odot$ is known, the distance can be estimated through the
expression $L_*=4\pi F_{\rm phot} d^²$ where $F_{\rm phot}$ is the
stellar flux observed and can be found by integrating the dereddened
photospheric model that was fitted to the photometry.

\begin{center}
\begin{figure}
\includegraphics[angle=0,scale=0.9]{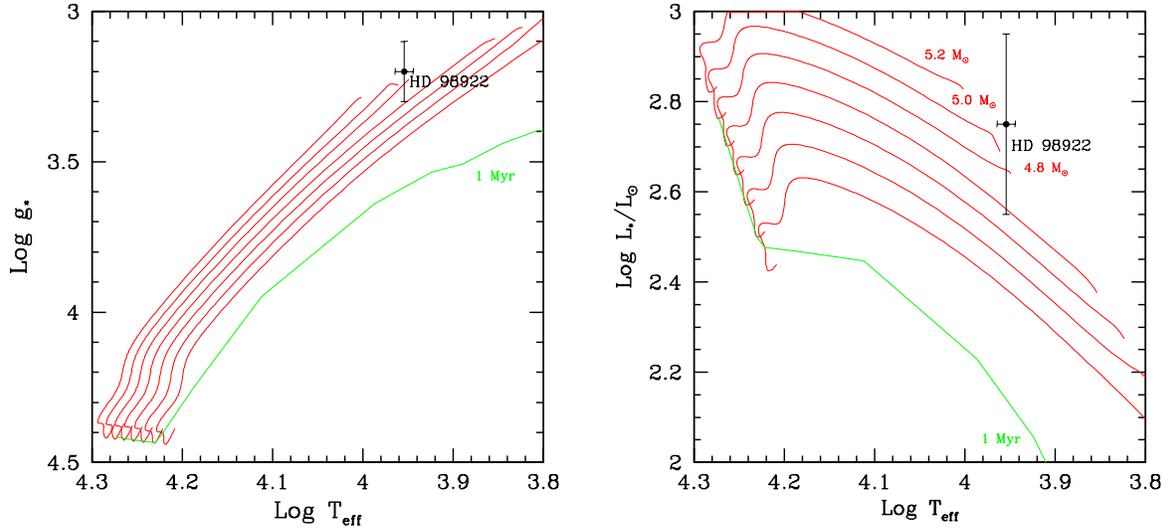}
\caption
{$\log g_* - \log T_{\rm eff}$ and $\log L_*/L_\odot - \log T_{\rm
    eff}$ HR diagrams showing evolutionary tracks for 4.0, 4.2, 4.4,
  4.6, 4.8, 5.0 and 5.2 $M_\odot$ and the 1-Myr isochrone from the
  Yonsei-Yale collection for $Z=0.007$. The position of a star with
  $T_{\rm eff}=9000$ K and $\log g_*=3.2$ in the diagram on the left
  is translated in a one-to-one correspondence to the HR diagram on
  the right to obtain the luminosity.  The value of $L_*$, and the
  estimate of the dereddened observed photospheric flux, $F_{\rm
    phot}$ allows us to calculate the distance to the star (see text
  for details). The position of star with respect to the isochrone
  indicates that the star is younger than 1 Myr.}
\label{HD98922_HR}
\end{figure}
\end{center}

In the panel on the left hand side of Fig. \ref{HD98922_HR} an HR
diagram $\log g_*$ -- $\log T_{\rm eff}$ with the PMS evolutionary
tracks from the Yonsei-Yale (Yi et al.  2001) collection for $Z=0.007$
--which corresponds roughly to [Fe/H]=$-0.5$-- has been plotted. The
tracks correspond to 4.0, 4.2, 4.4, 4.6, 4.8, 5.0 and 5.2
$M_\odot$. The point (9000~K, 3.0) is higher than the beginning of the
tracks, so, to carry out the estimate we have used instead a slightly
higher value of $\log g_*$, namely 3.20. The point (9000~K, $3.20\pm
0.1$) translates to the HR diagram $\log L_*/L_\odot$ -- $\log T_{\rm
  eff}$, on the right hand side of the figure, to a point with $\log
L_*/L_\odot=2.75\pm0.20$, therefore, the luminosity of the star would
be $L_*/L_\odot=562^{+329}_{-208}$.  The distance implied is
$d=507^{+131}_{104}$ pc. From this exercise we can also estimate the
mass (around 5 $M_\odot$), whereas the age seems to be less than 1
Myr.

We must note that the parameters we give in this paper have been
estimated under the hypothesis that this star is a single object.  A
binary scenario has also been proposed for HD~98922
\cite[e.g.][]{2006A&A...456.1045B}, which could potentially explain
the observed SED, the ultraviolet spectrum, and would place the system
further away (beyond 600~pc). This and other possible scenarios for
the stellar nature of the HD~98922 system will be deferred to a future
publication.

Similarly to the modeling presented for HD~104237
(Section~\ref{rthd104237}), we have used the MCFOST radiative transfer
code (Pinte et al. 2006) to reproduce both the observed SED and the
$^{12}$CO(3-2) emission the HD~98922 disk.  Photometric data available
from the literature is listed in Table~\ref{tablefluxes2}, together with the PACS
photometry was incorporated to our SED
fitting. Table~\ref{tablemodel2} shows the disk parameters that
reproduce the SED and the CO data.

%
%
%
%
%

We find that the SED and CO emission can be fit with dust disk mass
$2\times10^{-5}\,$M$_{\odot}$, of 320~AU in radius. Similarly to
HD~104237, no flaring with increasing radius is required to fit the
SED. No CO freezout is required to fit the CO spectrum. Since the
infrared emission is optically thick, the lack of longer wavelength
data makes very difficult to estimate the dust mass. 
The $^{12}$CO(3-2) data alone cannot be used to constrain the total
disk mass. It is found to be optically thick in disks, hence tracing
the surface temperature and not the underlying density field near the
mid plane. Moreover, it depends on many variable parameters such as
the assumption of T$_{gas}=$T$_{dust}$, CO freezout temperatures, the
dust to gas ratio, and the CO abundance with respect to H$_2$. Longer
wavelength follow-up (e.g. sub-mm) is required to further constrain
the mass of the disk.\\



%
%

\section{Discussion}\label{disc}

\begin{center}
\begin{figure}
\includegraphics[angle=0,scale=0.9]{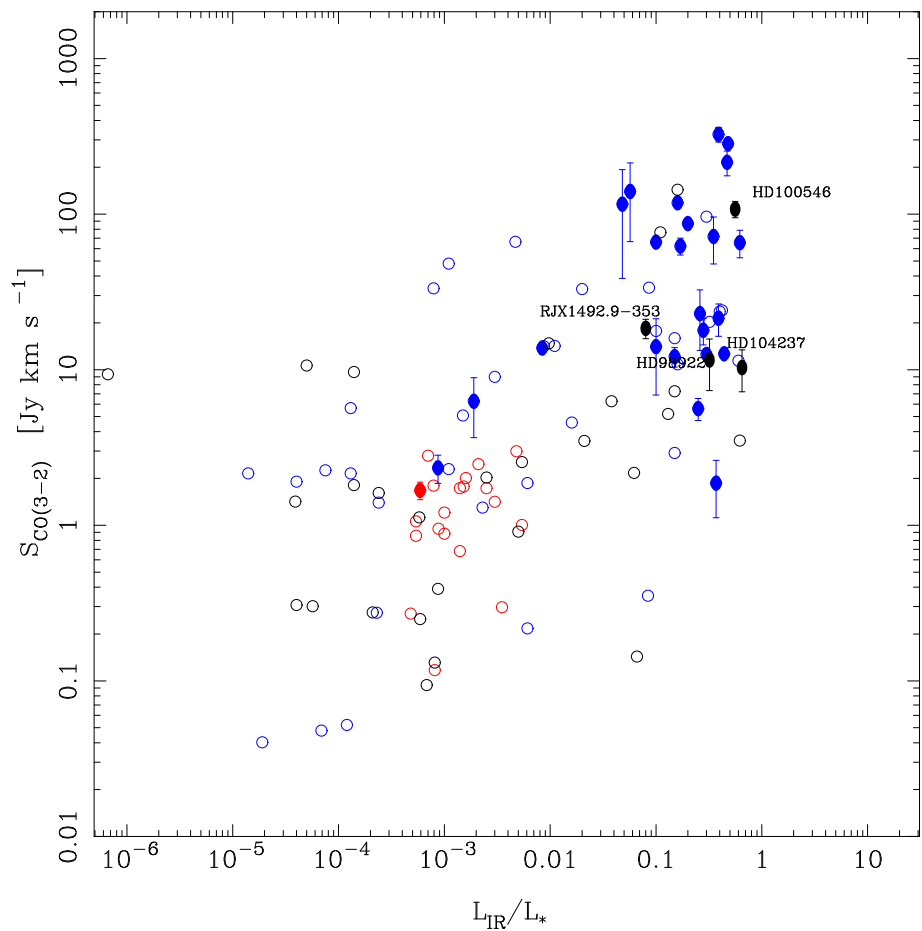}
\caption
{ Integrated CO intensity or upper limits versus fractional infrared
  excess for the observed stars (only stars with no cloud
  contamination, and having both distance estimates and enough SED
  points for reliable determinations have been included). Black filled
  circles denote objects with positive detections likely to be
  associated to circumstellar CO disks, while black open circles are
  1-sigma upper limits (computed using 10 km s$^{-1}$ windows, see
  text). Data from the literature for protoplanetary and debris disk
  from the CO surveys of \citet{dent05} and \citet{moor2011} are
  plotted in blue and red respectively.  }
\label{figcoir}
\end{figure}
\end{center}

We surveyed a sample of 52 stars covering a broad range of $L_{\rm IR
}/L_{*}$ values. We detect CO(3-2) emission from 20 systems, and 18
(90\%) of these have $L_{\rm IR }/L_{*} > 0.01$. This implies that in
the 31 systems with $L_{\rm IR }/L_{*} > 0.01$, we obtain a
detectability rate of 58\%. Figure~\ref{figcoir} shows the integrated
CO intensity versus fractional infrared excess for our targets. We
have not included the sources with possible background/foreground
cloud contamination, and also removed HD~155448.  The CO intensities
have been normalized to a distance of 100~pc\footnote{The conversion
  factors used to convert from Kelvin to Jansky were: 78.3 for ASTE
  (from T$^{*}_{A}$), 40.6 for APEX (from T$^{*}_{A}$), 18.5 for the
  JCMT (from T$_{\tt mb}$, since those are the units listed in
  \citet{dent05}, conversion factor obtained from \tt
  http://docs.jach.hawaii.edu/JCMT/HET/GUIDE/het\_guide.ps).} . Our
results are in agreement with the previous northern survey of
\citet{dent05}, where $\sim50\%$ of the stars $L_{\rm
  IR}$/$L_{*}>0.01$ have CO detections.

Ten objects (42\%) deviate from that trend by having $L_{\rm
  IR}$/$L_{*}>0.01$ but no detectable CO emission. HD~100453 and
HD~95881 have indeed been identified as disks in transition from
gas-rich to gas-poor \citep{collins2009,verhoeff2010}. HD~95881
appears to be in a particular transition between a gas-rich flaring
disk to gas-poor self-shadowed disk \citep{verhoeff2010}, which would
explain the relatively large infrared excess and the lack of
detectable CO. Overall, the results from this and previous surveys
indicate that the infrared excess is an indicator for detecting CO
gas, in agreement with previous studies that suggest that young stars
with large amounts of circumstellar dust emission also hold large
reservoirs of circumstellar molecular gas \citep{dent05}.

Only four of the newly detected systems are, however, clearly free of
CO emission from extended gas associated with ambient molecular cloud
material.  The contaminated sources may indeed harbor significant
quantities of CO, but higher resolution interferometric observation
would be needed to filter out any cloud contamination from the disk
emission. Alternatively, higher density tracers such as HCO$^+$ or CN
could be considered, as \citep{guilloteau2013,degregoriomonsalvo}.
High density tracers will detect the emission from the disk, while the
possible contaminating clouds are not dense enough to be detected in
these molecules. This is one of the main results from the pilot
programme from \citet{guilloteau2013}. In addition, higher order
transition in CO and HCO+ can also be used to avoid contamination from
cold clouds, as the temperatures required to excite these transitions
are largar than usually found in molecular clouds (see
Section~\ref{achamp}).

Most of the sources showing cloud contamination are all well studied
  disks known to harbor gas emission (e.g. HD~250550, HD~97048,
  HD~142527, HD~144668, HD~149914, KK~Oph and VV Ser). Our single-dish
  observations indicate they are still associated to their parent
  clouds, given the coincidence in position and velocity between the
  observed spectra and existing molecular cloud maps.

We detect $^{12}$CO around HD~104237 and HD~98922 for the first time,
and combine it with SED data to produce disk models. We confirm that
geometrically flat disk models can indeed reproduce both the SEDs and
CO profiles. The incorporation of CO freezout is required to account
for observed CO intensities of HD~104237, and is expected to play a
significant role if according to new models the CO is located closer
to the midplane where the temperatures are lower.  The 63.2 $\mu$m
[O~I] and the [O I] 6300 {\AA} line are also detected in both
disks. \citet{meeus12} found a strong correlation between 63.2 $\mu$m
[O~I] and $^{12}$CO(3-2) in a sample of 20 Herbig Ae/Be stars, but no
clear correlation between $^{12}$CO and [O I] 6300 {\AA}. Even though
the origin of the [O I] 6300 {\AA} line is unclear, it is usually
thought to originate in the surface of the disks \citep{acke05}, and
to trace the inner regions of the disk
\citep[1-50~AU;][]{2008A&A...485..487V}. The fact we are indeed
detecting both [O I] lines and the $^{12}$CO(3-2) may indicate that
the $^{12}$CO(3-2) emission stems from the warmer layers of the disk,
although these lines comes from different radii ranges with CO(3-2)
being form the most outside \citep[R$\ge\,$100~AU;
  e.g.][]{2008ApJ...678.1119H}.


HD~104237 is part of a multiple system, and our LABOCA imaging shows
that at least two sources in the field of view have detectable sub-mm
emission. Our SED modelling indicate a total disk mass of 
$4\times10^{-2}$~M$_{\odot}$ around the primary component. We can also
use the 870~$\mu$m flux to estimate the mass of the disk around
component E (which may also include a contribution from component
D). Assuming T$\sim$20~K, opacity of 0.35 m$^2/$Kg and 100:1
gas-to-dust \citep[e.g.,][]{2005ApJ...631.1134A}, implies a
0.004~M$_\odot$ disk mass for component E(+D).

The HD~104237 system bears some striking resemblance to
RW~Aur. RW~Aur~A is a very active classical T Tauri driving a powerful
optical jet.  It is separated by 1.2" from RW~Aur B, and there is also
some suspicion that RW~Aur~A may have a close companion (from radial
velocity variations). Radio interferometric observations of the
$^{12}$CO and $^{13}$CO lines in RW~Aur with the Plateau de Bure
interferometer \citep{2006A&A...452..897C} show that the RW~Aur~A disk
is very compact (R$_{out}$ $\sim$ 40$-$57 AU) probably due to truncation by
RW~Aur-B.  Despite the truncation, a large scale spiral arm extending
over 500~AU was detected in $^{12}$CO connecting the RW~Aur~A and
RW~Aur B disks, interpreted as tidal disruption of the RW~Aur-A disk
by the companion, to be searched for in the HD~104237 multiple system.

Based on the 1.26'' separation between HD~104237A and HD~104237B,
\citep{grady04} predicted a tidal truncation outer radius of 38~AU for
HD~104237A. This is significantly smaller than the 90~AU
characteristic radius we derive from the SED modelling. However the
truncation radius prediction is subject to uncertainties on the projected
distance, while the outer radius disk estimates is known to be poorly
constrained by pure SED fitting. Resolved imagining is required to 
directly measure the outer truncation radius of the disk.

\section{Conclusions}\label{conc}

We conducted a single-dish survey for $^{12}$CO(3-2) gas in a sample
of 52 protoplanetary and debris disks. We report a 58$\%$ detection
rate in sources with infrared excess fraction larger than 0.01. Most
of those detections can, however, be associated with a parent
molecular cloud, and therefore can only be considered upper limits to
the disk emission. We identify four CO detections that can be
explained as arising from circumstellar CO gas. We present the
detection of two new gas-rich disks, HD~104237 and HD~98922, and use
radiative transfer codes to model their continuum and gas
emission. Both disks are unusual, geometrically-flat, CO-rich disks
with 63.2 $\mu$m [O~I] and [O I] 6300 {\AA} line emission.  HD~104237
and HD~98922 are unique laboratories to study the extreme of
SED-predicted flat disks. Are the SEDs of `flat' disks different to
those of `flared' disks because of dust settling into the mid-plane or
are they just smaller in radii?  Resolved sub-mm images with existing
interferometers (e.g. ALMA) are required to further constrain the
physico-chemico structure of these disks and ultimately understand the
gas-clearing processes which yield planet formation.


\section*{Acknowledgments}

We are grateful to Mario van den Ancker and Andre M\"uller for
providing the UVES/VLT spectrum of HD 98922 and for fruitful
discussions on the determination of the parameters for this star.
Based on observations carried out with the Atacama Pathfinder
Experiment telescope (APEX). APEX is a collaboration between the
Max-Planck Institut fr Radioastronomie, the European Southern
Observatory, and the Onsala Space Observatory.  AH, SC, FM, SP and
WFRD acknowledge support from Millennium Science Initiative, Chilean
Ministry of Economy: Nucleus P10-022-F.  AP and JMT acknowledge
support from MICINN (Spain) AYA2011-30228-C03 grants (co-funded with
FEDER funds) AGAUR (Catalonia) 2009SGR1172 grant. The ICC (UB) is a
CSIC-Associated Unit through the ICE (CSIC). Finally, we would like to
thank our anonymous referee for a very careful and detailed review
which we think has improved very significantly our paper.

{}

\clearpage

\begin{landscape}

\begin{deluxetable}{lllcccccll} \label{table1}
\tabletypesize{\scriptsize}
\tablewidth{0pt}
 \setlength{\tabcolsep}{1mm}
\tablecaption{Target Sample}
\tablehead{
\colhead{} & \colhead{Star} & \colhead{Ra} & \colhead{Dec} & \colhead{Sp.} &  \colhead{Distance}  & \colhead{$L_{\rm IR}/L_{*}$} &  
\colhead{CO intensity$^f$ }  & \colhead{Disk group} & \colhead{Reference}\\
\colhead{} &\colhead{} & \colhead{(J2000)}  & \colhead{(J2000)} & \colhead{Type} & 
\colhead{ (pc)}  & \colhead{} & 
\colhead{(K~km~s$^{-1}$)} & \colhead{}& \colhead{}\\
}
\startdata
1&HD~105         &00:05:52.5 &	-41:45:11.0   & G0V   &  40    & 5.9$\times 10^{-4}$  & $<0.02^e$                & Debris Disk & \citet{hillenbrand08}  \\
2&HR~10	         &00:07:18.2 &	-17:23:13.1   &A2IV/V & 160    & 6.6$\times 10^{-7}$  & $<0.09^c$                & Debris Disk & Possible bogus IRAS identification \citep{fajardo98a}. \\
3&HD~377	 &00:08:25.7 &	+06:37:00.4   & G2V   &  40    & 5.8$\times 10^{-4}$  & $<0.09^a$                & Debris Disk & \citet{hillenbrand08} \\
4&HD~3003        &00:32:43.9 &	-63:01:53.4   & A0V   &  46    & 1.4$\times 10^{-4}$  & $<0.11^a$                & Debris Disk & \citet{rhee07a} \\
5&HD~12039       &01:57:48.9 &	-21:54:05.3   & G4V   &  42    & 2.1$\times 10^{-4}$  & $<0.02^e$                & Debris Disk & \citet{rhee07a} \\
6&HD~17848       &02:49:01.4 &	-62:48:23.4   & A2V   &  50    & 3.9$\times 10^{-5}$  & $<0.14^d$                & Debris Disk & \citet{sylvester00}\\
7&HD~21563       &03:24:02.3 &	-69:37:28.5   & A4V   & 182    & 9.7$\times 10^{-3}$  & $<0.11^d$                & Debris Disk & \citet{sylvester00} \\
8&HD~28001       &04:22:05.9 &	-56:58:59.5   & A4V   &  -     & 5.0$\times 10^{-3}$  & $<0.11^d$                &  Debris Disk& \citet{sylvester00} \\
9&HD~32297       &05:02:27.4 &	+07:27:39.6   & A0    & 112    & 5.4$\times 10^{-3}$  & $<0.05^c$       	& Debris Disk& \citet{rhee07a}; $L_{\rm IR}/L_{*}$ from \citet{meeus12} \\
10&HD~36917      &05:34:46.9 &	-05:34:14.5   & A0V   & 460(5) & 1.0$\times 10^{-1}$  & 1.1 $\pm$~0.5$^e$        & HAeBe disk& \citet{malfait98}\\
11&HD~37411      &05:38:14.5 &	-05:25:13.2   &B9 V   & 510(13)& 3.9$\times 10^{-1}$  & 0.5 $\pm$ 0.25$^e$    & HAeBe disk& \citet{malfait98} \\
12&HD~37258      &05:36:59.2 &	-06:09:16.3   & A2V   & 510(6) & 3.7$\times 10^{-1}$  & 4.8 $\pm$~1.2$^b$        & HAeBe disk& \citet{malfait98} \\
13&HD~37389      &05:38:08.0 &	-01:45:07.8   & A0    & 210(7) & 1.2$\times 10^{-2}$  & 3.1 $\pm$~0.5$^b$        & HAeBe disk&\citet{coulson98} \\
14&HD~38087      &05:43:00.5 &	-02:18:45.3   & B5 V  & 199    & 1.9$\times 10^{-2}$  & 4.5 $\pm$~1.1$^e$        & HAeBe disk& \citet{malfait98}  \\
15&HD~39014      &05:44:46.3 &	-65:44:07.9   & A7 V  &  44    & 5.7$\times 10^{-5}$  & $<0.02^e$                & HAeBe disk& \citet{malfait98}  \\
16&HD~250550     &06:01:59.9 &	+16:30:56.7   & B9    & 606(8) & 1.4                  &7.9 $\pm$~1.7$^c$        & HAeBe disk& \citet{dewinter01} \\
17&HD~61005      &07:35:47.4 &	-32:12:14.0   & G8V   &  34    & 6.8$\times 10^{-4}$  & $<$0.02$^c$              & Debris Disk& \citet{hillenbrand08} \\
18&HD~85567      &09:50:28.5 &	-60:58:02.9   & B2    & 770(9) & 3.3$\times 10^{-1}$  &1.4 $\pm$~0.4$^e$         & HAeBe disk& \citet{malfait98}  \\
19&HD~95881      &11:01:57.6 &	-71:30:48.3   &A1/2III& 118(1) & 6.2$\times 10^{-2}$  & $<$0.02$^e$              & HAeBe disk& \citet{malfait98}  \\
20&HD~97048      &11:08:03.3 &	-77:39:17.4   & A0pshe& 175    & 3.9$\times 10^{-1}$  &1.4 $\pm$~0.4$^c$    & HAeBe disk& \citet{dewinter01}; $L_{\rm IR}/L_{*}$ from \citet{meeus12} \\
21&Hen3-600      &11:10:27.8 &	-37:31:52.0   & M4Ve  &  42(2) & 6.6$\times 10^{-2}$  & $<$0.02$^c$              & T~Tauri Disk& \citet{rhee07b} \\
22&HD~98922      &11:22:31.6 &	-53:22:11.4   &A2 IIIe  & 115    & 6.5$\times 10^{-1}$  & \bf{0.11$ \pm$~0.03}$^e$ & HAeBe disk& \citet{malfait98}  \\
23&HD~100453     &11:33:05.5 &	-54:19:28.5   & A9Ve  & 111    & 6.2$\times 10^{-1}$  & $<$0.07$^b$              & HAeBe disk& \citet{malfait98}; $L_{\rm IR}/L_{*}$ from \citet{meeus12}  \\
24&HD~100546     &11:33:25.4 &	-70:11:41.2   &B9Vne  & 103    & 5.6$\times 10^{-1}$  &2.5 $\pm$~0.3$^c$    & HAeBe disk& \citet{malfait98}; $L_{\rm IR}/L_{*}$ from \citet{meeus12}  \\
25&HD~101412     &11:39:44.4 &	-60:10:27.7   &B9.5V  & 160(12)& 1.5$\times 10^{-1}$  & $<$0.07$^c$              & HAeBe disk& \citet{dewinter01} \\
26&HD~104237     &12:00:05.0 &	-78:11:34.5   &A4/5V& 116    & 3.2$\times 10^{-1}$    &\bf{0.11 $\pm$~0.04}$^e$  & HAeBe disk& \citet{malfait98}; $L_{\rm IR}/L_{*}$ from \citet{meeus12}  \\
27&HD~109573A    &12:36:01.0 &   -39:52:10.2  &A0     &  67    & 5.0$\times 10^{-3}$  & $<$0.05$^b$      	& Debris Disk& \citet{rhee07a}  \\
28&HD~110058     &12:39:46.1 &	-49:11:55.5   & A0V   & 100    & 2.5$\times 10^{-3}$  & $<$0.05$^c$              & Debris Disk& \citet{rhee07a} \\
29&HD~113766     &13:06:35.8 &	-46:02:02.0   & F4V   & 131    & 2.1$\times 10^{-2}$  & $<$0.05$^c$              & Debris Disk& \citet{rhee07b} \\
30&PDS~66	 &13:22:07.5 &	-69:38:12.1   & K1Ve  &  86(2) & 1.3$\times 10^{-1}$  & $<$0.09$^a$              & T~Tauri Disk& \citet{hillenbrand08} \\
31&HD~124237     &14:14:33.6 &	-61:47:56.3   &B5 V   & 452    & 1.6$\times 10^{-1}$  & $<0.09^e$                & HAeBe disk& \citet{malfait98}  \\
32&HD~132947     &15:04:56.0 &	-63:07:52.6   & A0    &  -     & 2.3$\times 10^{-2}$  & 0.26 $\pm$~0.05$^b$      & HAeBe disk& \citet{malfait98}  \\
33&HD~140863     &15:48:49.4 &	-57:37:55.3   &A0 III &  -     & 5.2$\times 10^{-1}$  & $<0.02^e$                & HAeBe disk& \citet{malfait98} \\
34&HD~142527     &15:56:41.8 &	-42:19:23.2   &F6IIIe & 198    & 9.8$\times 10^{-1}$  & 1.7 $\pm$~0.7$^b$        & HAeBe disk& \citet{malfait98}; $L_{\rm IR}/L_{*}$ from \citet{meeus12}  \\
35&ScoPMS31      &16:06:21.9 &	-19:28:44.5   & M0    & 145(10)& 1.2$\times 10^{-1}$  & $<0.02^e$                & T Tauri disk &\citet{dahm09} \\
36&HD~144668     &16:08:34.2 &	-39:06:18.3   &A7IVe  & 207    & 5.1$\times 10^{-1}$  & 7.2 $\pm$~2.0$^b$        & HAeBe disk& \citet{malfait98}; $L_{\rm IR}/L_{*}$ from \citet{meeus12} \\
37&HD~149914     &16:38:28.6 &	-18:13:13.7   &B9.5IV & 165    & 1.3$\times 10^{-3}$  & 1.1 $\pm$~0.4$^b$        & HAeBe disk& \citet{malfait98}  \\
38&HD~155448     &17:12:58.7 &	-32:14:33.5   & B 9   & 606    & 6.0$\times 10^{-2}$  &\bf{0.11 $\pm$~0.02}$^e$  & HAeBe disk& \citet{malfait98}  \\ 
39&KK~Oph        &17:10:08.0 &	-27:15:18.2   &A8Ve   & 160(4) & 2.0                 & 0.6 $\pm$~0.2$^a$        &  HAeBe disk& \citet{dewinter01}; $L_{\rm IR}/L_{*}$ from \citet{meeus12} \\
40&HD~158352     &17:28:49.6 &	+00:19:50.2   & A8Vp  &  63    & 1.6$\times 10^{-4}$  & $<$0.6$^c$               & Debris Disk& \citet{rhee07a}; $L_{\rm IR}/L_{*}$ from \citet{meeus12}  \\
41&HD~158643     &17:31:24.9 &	-23:57:45.5   & A0V   & 131    & 3.8$\times 10^{-2}$  & $<$0.09$^c$              & HAeBe disk& \citet{malfait98}  \\
42&sao185668     &17:43:55.6 &	-22:05:44.7   & B3    & 700(7) & 1.1$\times 10^{-1}$  & $<0.02^e$                & HAeBe disk& \citet{malfait98} \\
43&HR~6629	 &17:47:53.5 &	+02:42:26.2   & A0V   &  29    & 4.0$\times 10^{-5}$  & $<$0.09$^b$              & Debris Disk& \citet{fajardo98b}\\
44&HR~6723	 &18:01:45.1 &	+01:18:18.2   & A2Vn  &  81    & 5.0$\times 10^{-5}$  & $<$0.4$^c$               & Debris Disk& \citet{rieke05}   \\
45&VV~Ser	 &18:28:47.8 &  +00:08:39.9   & -     & 260(3) & 9.1                  &0.47 $\pm$~0.08$^c$      & HAeBe disk& \citet{dewinter01} \\
46&RX~J1842.9-353&18:42:57.9 &  -35:32:42.7   & K2    & 130    & 8.0$\times 10^{-2}$  &\bf{0.14 $\pm$~0.03}$^a$  & T~Tauri Disk& \citet{hillenbrand08} \\
47&HD~172555     &18:45:26.9 &  -64:52:16.5   & A7V   &  29    & 8.1$\times 10^{-4}$  & $<0.02^e$                & Debris Disk& \citet{rhee07a}  \\
48&RX~J1852.3-370&18:52:17.2 &  -37:00:11.9   &K7V    & 130    &  1.8$\times 10^{-1}$ & $<$0.09$^a$              & T~Tauri Disk& \citet{hillenbrand08} \\
49&HD~181327     &19:22:58.9 &	-54:32:16.9   & F6V   &  50    & 8.7$\times 10^{-4}$  & $<0.02^e$                & Debris Disk& \citet{rhee07a}  \\
50&HD~178253     &19:09:28.3 &	-37:54:16.1   & A2V   &  39    & 3.9$\times 10^{-5}$  & 0.57 $\pm$~0.1$^e$       & Debris Disk& \citet{sylvester00} \\
51&HD~181296     &19:22:51.2 &	-54:25:26.1   & A0V   &  47    & 2.4$\times 10^{-4}$  & $<0.18^d$                & Debris Disk& \citet{rhee07a}  \\
52&HD~184800     &19:38:40.8 &	-51:00:18.5   &A8/9V  &  -     & 1.3$\times 10^{-1}$  & $<0.09^b$                & Debris Disk& \citet{sylvester00}  \\
%
\enddata \tablenotetext{a}{ASTE, June 2008} \tablenotetext{b}{APEX,
  June 2008} \tablenotetext{c}{APEX, July 2009}
\tablenotetext{d}{APEX, December 2009} \tablenotetext{e}{ASTE, July
  2010} \tablenotetext{f}{Intensity units are in antenna temperatures
  (T$^{*}_{A}$). Upper limits were computed over 10~km~s$^{-1}$
  velocity interval. }

Spectral types are from Michigan SpectralCatalogue and distances are
derived from parallax measures listed in the {\em Hipparcos}
catalogue; other references for distance are: 1) \citet{acke04}, 2)
\citet{rhee07b}, 3) \citet{straizys96}, 4) \citet{sandell2011} - 5)
Distance to the Ori OB1 association e.g. \cite{hamaguchi05}, (6)
\citet{liu11}, 7) \citet{coulson98}, 8) \citet{brittain07}, 9)
\citet{vieira03}, 10) Distance to the Upper Scorpius OB association
e.g. \cite{dezeeuw99}, 11) \citet{kun00}, 12) \citet{acke05}, 13)
\citet{wade07}. The four detections free of CO emission from
extended gas associated with ambient molecular cloud material are
marked in bold faces.

\end{deluxetable}


%
\label{table1}

\end{landscape}

\clearpage

\begin{table}
\caption{HD~104237  photometric data}
\label{tablefluxes1}
\begin{center}
\leavevmode
\begin{tabular}{ccccl} \hline \hline    
Wavelength    &      Flux             &  Flux Error             & Reference & Reference   \\  
($\mu$m)     & (erg/cm$^{-2}$/s/\AA) &  (erg/cm$^{-2}$/s/\AA)  & Filter     &  \\  
\hline \hline 
   0.345  &6.8022E-12 &  6.8022E-13   & u          &  \citet{hauck1998} \\
   0.347  &6.9823E-12 &  6.9823E-13   & U          &  \citet{malfait98}    \\
   0.411  &1.3423E-11 &  1.3423E-12   & v          &  \citet{hauck1998}  \\
   0.425  &1.2706E-11 &  1.2706E-12   & B          &  \citet{malfait98}    \\
   0.440  &1.2443E-11 &  1.6045E-13   & B          &  \citet{hog2000}  \\
   0.467  &1.2078E-11 &  1.2078E-12   & b          &  \citet{hauck1998} \\
   0.548  &8.5312E-12 &  8.5312E-13   & y          &  \citet{hauck1998}  \\
   0.550  &8.7983E-12 &  8.7983E-13   & V          &  \citet{malfait98}    \\
   0.550  &8.3176E-12 &  7.6609E-14   & V          &  \citet{hog2000}    \\
   0.640  &5.1643E-12 &  5.1643E-13   & R          &  \citet{2003yCat.2246....0C}  \\
   1.215  &1.6590E-12 &  1.6590E-13   & J          &  \citet{malfait98}   \\
   1.235  &1.4818E-12 &  3.1391E-14   & J          &  \citet{2003yCat.2246....0C}  \\
   1.654  &1.0114E-12 &  1.0114E-13   & H          &  \citet{malfait98}   \\
   1.662  &8.8605E-13 &  4.8173E-14   & H          &  \citet{2003yCat.2246....0C}  \\
   2.159  &6.2851E-13 &  1.0420E-14   & K$_s$      &  \citet{2003yCat.2246....0C} \\
   2.179  &7.0558E-13 &  7.0558E-14   & K          &  \citet{malfait98}    \\
   3.547  &3.9628E-13 &  3.9628E-14   & L          &  \citet{malfait98}    \\
   4.769  &1.9592E-13 &  1.9592E-14   & M          &  \citet{malfait98}    \\
  12.000  &3.4073E-14 &  3.4073E-15   & 12~$\mu$m  & \citet{1988iras....1.....B} \\
  25.000  &7.8464E-15 &  7.8464E-16   & 25~$\mu$m  & \citet{1988iras....1.....B} \\
  60.000  &7.0254E-16 &  7.0254E-17   & 60~$\mu$m  & \citet{1988iras....1.....B} \\
\hline   
    \end{tabular}
  \end{center}
\tablecomments{Table 2 is published in its entirety in the electronic
edition of the Journal.  A portion is shown here for
guidance regarding its form and content.}
\end{table}

\clearpage

\begin{table}
  \caption{HD~104237 disk model}
  \label{tablemodel1}
  \begin{center}
    \leavevmode
    \begin{tabular}{lcl} \hline \hline              
   Physical parameter  & Value &  Reference   \\
 \hline    \hline          
  Stellar properties &   &  \\
\hline   
Spectral type 1                          &  A4V   & \citet{tatulli07}  \\
Spectral type 2                          &  K3V   &  \citet{tatulli07} \\
Primary star Mass:   M$_1$ (M$_{\odot}$)  &  2.2   &  \citet{garcia2013}  \\
Secondary star Mass :M$_2$ (M$_{\odot}$)  &  1.4   &  \citet{garcia2013}  \\
Effective temperature:            T$_1$~(K)  &  8000   &   \citet{tatulli07} \\
Effective temperature:            T$_2$~(K)  &  4750   &    \citet{tatulli07}\\
Stellar Luminosity:            L$_1$ (L$_{\odot}$) &  30   &   \citet{tatulli07} \\
Stellar Luminosity:            L$_2$ (L$_{\odot}$) &   3   &   \citet{tatulli07} \\
Distance: d (pc) &  115   &   \citet{tatulli07} \\

 \hline 
  Disk structure   &   &\\
 \hline 
Disk total mass: M$_D$ (M$_{\odot}$) & $100\times $M$_d$    & this work   \\
Disk dust mass M$_d$  (M$_{\odot}$) &  $4\times10^{-4}$     &  this work \\
Inner Rim: R$_{in}$ (AU)  & 0.45    &  \citet{tatulli07} \\
Characteristic radius   R$_{C}$ (AU) &  90   &  this work \\
Characteristic height   h$_{C}$ (AU) & 6    &  this work \\
Surface density exponent: $\gamma$ &   0.8 &  this work \\ 
Flaring exponent: $\psi$ &   0.0   &  this work \\ 
Inclination angle: i (deg)  &  31   &  this work \\
Systemic Velocity: $\rm{V}_{\rm{lsr}}$ (km/s) &  4.2   &  this work  \\ 
 \hline  
    \end{tabular}
  \end{center}
\end{table}

\clearpage

\begin{table}
\caption{HD~98922  photometric data}
\label{tablefluxes2}
\begin{center}
\leavevmode
\begin{tabular}{ccccl} \hline \hline    
Wavelength   &Flux                    &  Flux Error          & Reference   & Reference   \\  
($\mu$m)     & (erg/cm$^{-2}$/s/\AA) &  (erg/cm$^{-2}$/s/\AA)  & Filter     &  \\  
\hline \hline 
   0.347 &  6.7298E-12 &  6.7298E-13  &U  &  \citet{malfait98}  \\
   0.425 &  1.3305E-11 &  1.3305E-12  &B  &  \citet{malfait98}  \\
   0.440 &  1.2489E-11 &  1.2489E-12  &B  &  \citet{hog2000} \\
   0.550 &  7.3858E-12 &  7.3858E-13  &V &  \citet{malfait98}  \\
   0.550 &  7.5370E-12 &  7.5370E-13  &V  &  \citet{hog2000} \\
   0.548 &  7.1682E-12 &  3.3011E-13  &y  &  \citet{hauck1998} \\
   1.215 &  1.1478E-12 &  1.1478E-13  &J  &  \citet{malfait98}  \\
   1.235 &  1.2427E-12 &  2.2893E-14  &J  &  \citet{2003yCat.2246....0C}  \\
   1.654 &  8.2592E-13 &  8.2592E-14  &H  &  \citet{malfait98}  \\
   1.662 &  9.0252E-13 &  2.4109E-14  &H  &  \citet{2003yCat.2246....0C}  \\
   2.179 &  7.1211E-13 &  7.1211E-14  &K  &  \citet{malfait98}  \\
   2.159 &  8.3236E-13 &  2.7604E-14  &K$_s$    &  \citet{2003yCat.2246....0C}  \\
   3.547 &  4.9889E-13 &  4.9889E-14  &K  &  \citet{malfait98}  \\
   4.769 &  2.6797E-13 &  2.6797E-14  &M  &  \citet{malfait98}  \\
  12.000 &  6.3403E-14 &  6.3403E-15  &12~$\mu$m  &  \citet{1988iras....1.....B}  \\
  25.000 &  9.2532E-15 &  9.2532E-16  &25~$\mu$m  &  \citet{1988iras....1.....B}   \\
  60.000 &  3.5550E-16 &  3.5550E-17  &60~$\mu$m  &  \citet{1988iras....1.....B}    \\
  65.000 &  2.3016e-16 &  1.6985e-17  & 65~$\mu$m  &   \citet{yamamura2011}       \\
  70.000 &  2.1909e-16 &  1.0954e-17  & 70~$\mu$m  &   PACS (This work)       \\ 
  90.000 &  9.5968e-17 &  5.2938e-18  & 90~$\mu$m   &   \citet{yamamura2011}       \\
 140.000 &  1.7314e-17 &  7.9813e-18  & 140~$\mu$m  &   \citet{yamamura2011}       \\ 
 160.000 &  1.0270e-17 &  5.1351e-19   & 160~$\mu$m  &    PACS (This work)       \\ 
\hline   
    \end{tabular}
  \end{center}

\end{table}

\clearpage

\begin{table}
  \caption{HD~98922 disk model}
  \label{tablemodel2}
  \begin{center}
    \leavevmode
    \begin{tabular}{lcl} \hline \hline              
   Physical parameter  & Value &  Reference   \\
 \hline    \hline          
  Stellar properties &   &  \\
\hline   
Spectral type    &  A2~III   &  This work \\
Primary star Mass: M (M$_{\odot}$)  &  5.0  & This work  \\
Effective temperature: T ~(K) &  9000   &  This work \\
Stellar Radius: R (R$_{\odot}$) &  9.0   &  This work  \\
Distance: d (pc) &  $d=507$ &   This work \\
 \hline 
  Disk structure   &   &\\
 \hline 
Disk total mass: M$_D$ (M$_{\odot}$) &  $100\times $M$_d$    & this work  \\
Disk dust mass  M$_d$  (M$_{\odot}$)  &   $2\times10^{-5}$     &  this work \\
Inner Rim: R$_{in}$ (AU)  &  1.5   & this work  \\
Characteristic radius   R$_{C}$ (AU) &  320   & this work  \\
Characteristic height   h$_{C}$ (AU) &  0.15  &  this work \\
Surface density exponent: $\gamma$ &   1.0 &  this work \\ 
Flaring exponent: $\psi$ &   0.0  &  this work \\ 
Inclination angle: i (deg)  &  20   & this work  \\
Systemic Velocity: $\rm{V}_{\rm{lsr}}$ & -8    & this work  \\

 \hline  
    \end{tabular}
  \end{center}
\end{table}

\clearpage

\clearpage

\begin{center}
\begin{figure}
\includegraphics[angle=0,scale=0.85]{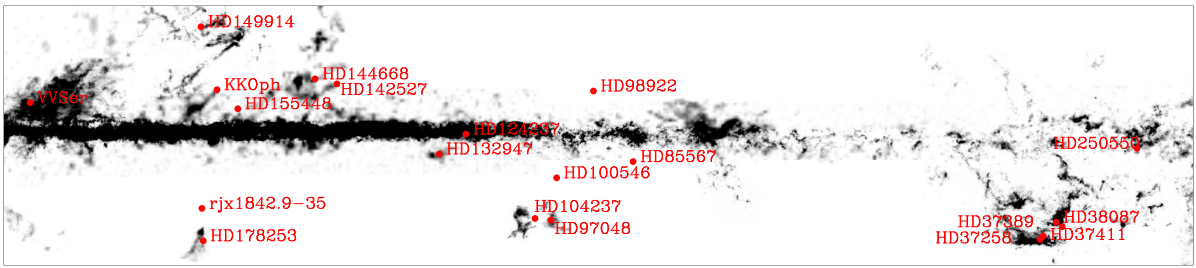}
\caption{Location of southern HAeBe with CO(3-2) detections shown against known molecular clouds CO maps (from Milky Way CO \citet{dame01}).}\label{figmilkyway}
\end{figure}
\end{center}

\clearpage

\begin{center}
\begin{figure}
\includegraphics[angle=0,scale=0.6]{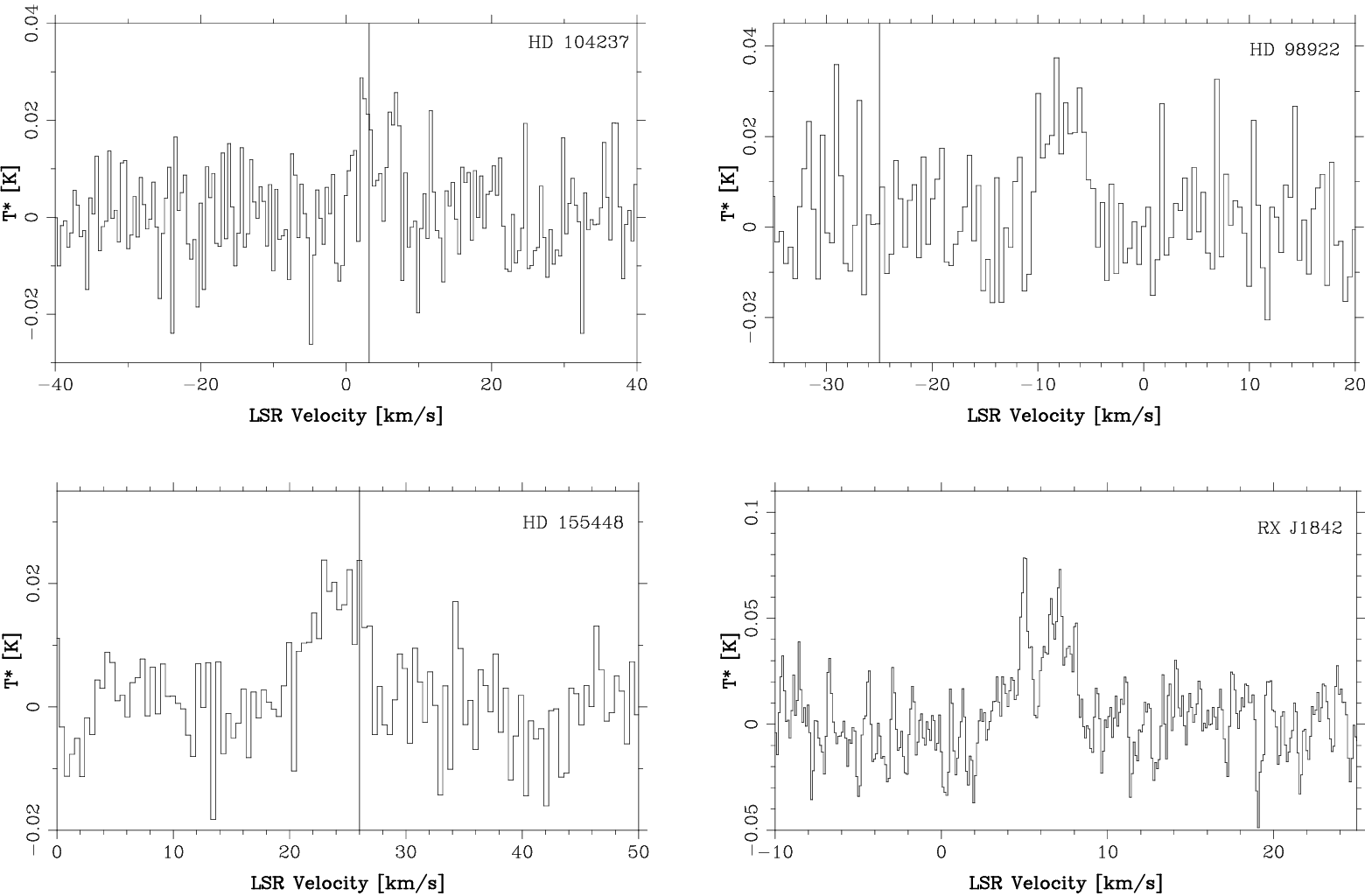}
\caption{ASTE spectra for the four new CO disk detections. When available, the stellar velocity has been plotted with a vertical line (in $\rm{V}_{\rm{lsr}}$). Intensity units are in antenna temperature (T$^{*}_{A}$).}\label{figco1}
\end{figure}\label{figco1}
\end{center}

\clearpage

\begin{center}
\begin{figure}
\includegraphics[angle=0,scale=1.2]{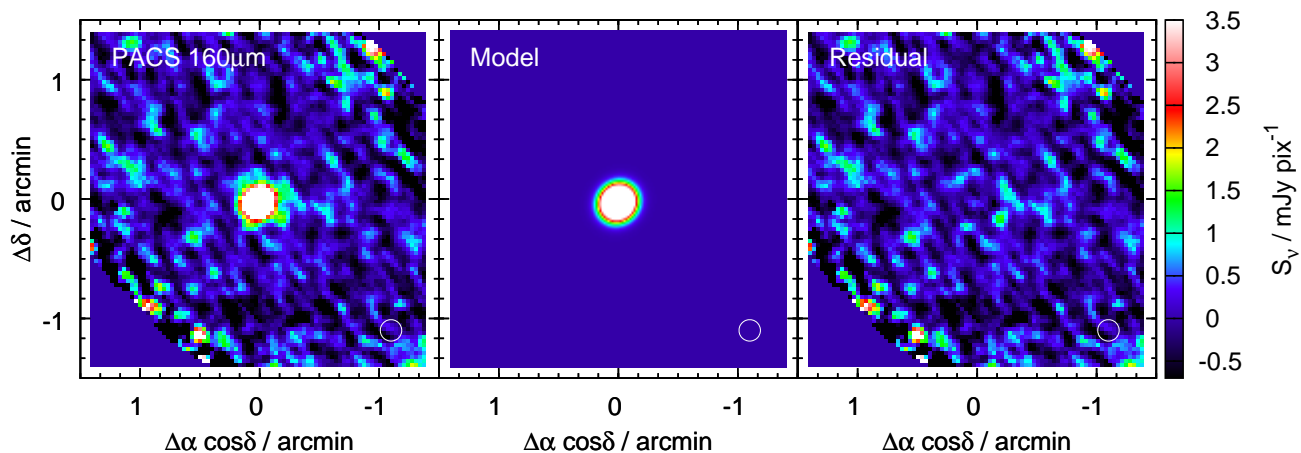}
\caption
{
 {\it{Left:}} {\it Herschel}/PACS $160\,\mu{\rm m}$ map around HD~98922 (from archive). {\it{Center:}} Gaussian source model used
for photometric extraction (total $F_\nu=878\pm44\,{\rm mJy}$, ${\rm
FWHM}=11.0\!\times\!11.9''$). {\it{Right:}} residual map after source
subtraction. FWHM beam circles of $10.7''$ diameter are shown.}
\label{pacs}
\end{figure}
\end{center}

%

\begin{center}
\begin{figure}
\includegraphics[angle=0,scale=0.5]{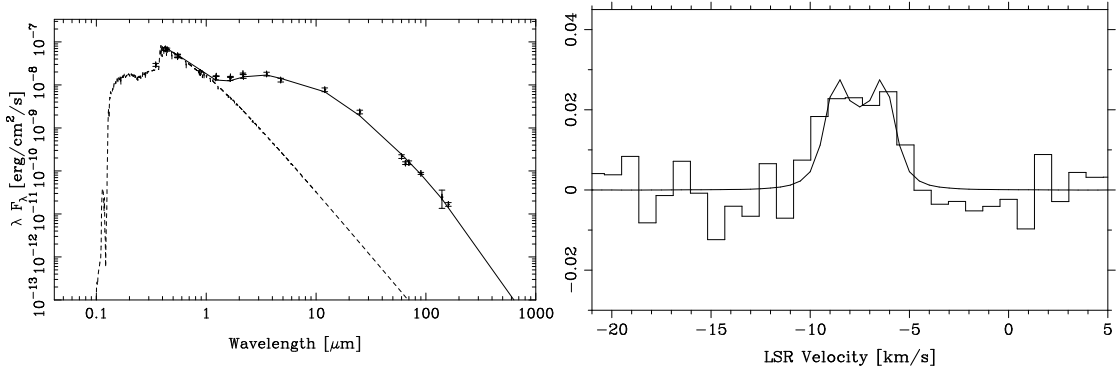}
\caption
               { {\it{Left:}} Observed SED of HD~98922 (black points
                 with error bars) plotted against the MCFOST disk
                 model that fits the optical to far-IR data (solid
                 line). The full photometric data is available in the
                 online version of the article. {\it{Right:}} Velocity
                 distribution of the $^{12}$CO(3-2) line detected from
                 the disk around HD~98922, with the disk model
                 superimposed (solid line). Intensity units are in
                 antenna temperature (T$^{*}_{A}$).  The spectrum
                 shown Figure~\ref{figco1} has been smoothed by
                 re-binning into 2 channel averages. }
\label{hd98922-sed}
\end{figure}
\end{center}

\clearpage

\appendix

\section{Appendix}\label{cont}

\subsection{Detections with background/foreground contamination}

In this section we present  individual sources  with CO detections
likely to arise from background/foreground contamination.

\subsubsection{HD~124237}

HD~124237 is located on the galactic plane \citep{dame01}, yielding
several velocity components of galactic CO detected in our spectra
(Figure~\ref{figco2}).  The line intensities are typical of galactic
clouds, and if present, any circumstellar CO emission would be
completely dominated by the ambient emission. The intensities of the
lines vary significantly when using a different reference position,
indication of a strong contamination from ambient CO.

\subsubsection{HD~178253}

This A2V star, alias $\alpha$~CrA, was originally catalogued as a
debris disk type star by \citet{mann98} based on data from
{\it{IRAS}}, and subsequently re-observed with {\it{Spitzer}} by
\citet{rieke05}. Mid-infrared imaging using Gemini South failed to
detect any significant mid-IR excess towards this star
\citep{moerchen2010}. \\

HD~178253 is located in the direction of the CrA molecular cloud
\citep{harju93}, which is clearly seen in the spectrum of the
reference (off-) position. HD~178253 has a photospheric radial velocity
of $\rm{V}_{\rm{helio}} = -18.4\pm2~$km~s$^{-1}$ \citep{barbier00}, or
$\rm{V}_{\rm{lsr}} = -12.3~$km~s$^{-1}$. No prominent emission is
detected near the tabulated stellar velocity.  In addition, as seen in
Figure~\ref{figco2}, the resulting spectrum changes dramatically when
using a different reference position. Therefore, the CO detection
cannot be unambiguously associated to a circumstellar disk.

\subsubsection{HD~36917, HD~37411, HD~37389, HD~37258 and  HD~38087}
 
\begin{center}
\begin{figure}
\includegraphics[angle=0,scale=0.6]{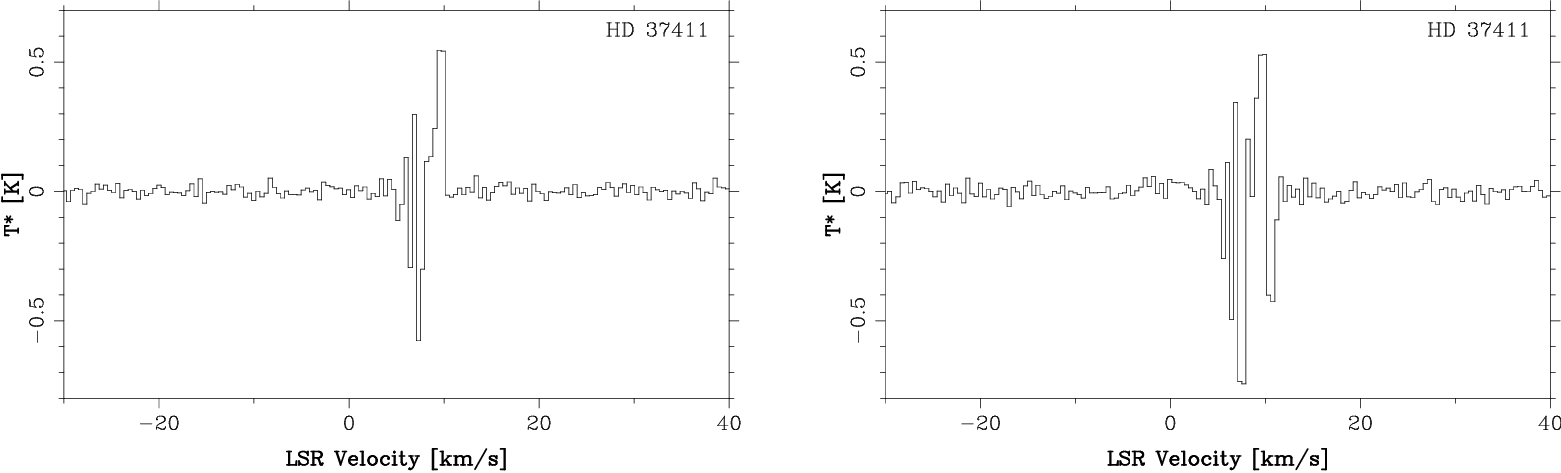}
\caption{Same as Figure~\ref{figco2}.}\label{figco3}
\end{figure}
\end{center}

HD~36917, HD~37411, HD~37389, HD~37258 and HD~38087 are located in the Orion
molecular clouds. The spectra toward these stars is contaminated due
to cloud emission in the reference. This is obvious in the case of
HD~36917 (Figure~\ref{figco2}), where a second observation with a
different reference position was acquired. Changing the off- position
yielded false absorption features indicative of strong emission in the
reference.  A similar effect is seen when comparing the two spectra
taken towards HD~37411 (Figure~\ref{figco3}). \\

Located in the middle of Orion's waist, between Alnilam and
$\zeta$~Ori, HD~37389 is surrounded by the strong background from
Orion B.  The single-pointing APEX CO(3-2) spectrum gives a narrow
line, with a velocity-integrated $T_\mathrm{mb}$ intensity of
4.2~K~km~s$^{-1}$ (Figure~\ref{figco4}). CO(3-2) probably stems from
the diffuse cloud.

Our CHAMP+ follow-up observations \citep{casassus13} show how the
diffuse emission is clearly traced in north-western horns of the
CHAMP+ footprint (see Appendix~\ref{achamp} for details).

For HD~37258 and HD~38087 the detected peaks appear too large compared
to pure disk emission (Figures~\ref{figco4} and \ref{figco5}).  Bogus
absorption features due to reference contamination are seen in both
spectra.  The smaller peaks at 7~km~s$^{-1}$ are too narrow to be
circumstellar, and also coincide with the 4-14 km s$^{-1}$ range of the
known clouds in Orion \citep{dame01}. CO emission toward these stars
is most likely associated to the Orion Molecular Cloud.

\subsubsection{HD85567 }

HD85567 is a B2Ve star first associated to the HAeBe membership by
\citet{the94}. Its infrared excess of 3.3$\times 10^{-1}$ is
characteristic of young, massive protoplanetary
disks. \citet{juhasz10} reports strong silicate emission detected with
the {{\it Spitzer Space Telescope}}

\begin{center}
\begin{figure}
\includegraphics[angle=0,scale=0.6]{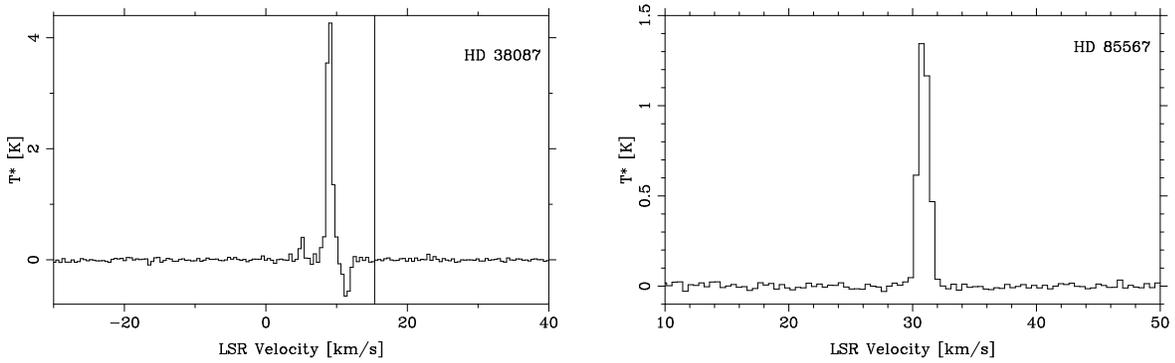}
\caption{Same as Figure~\ref{figco2}.  The stellar
velocity for HD~38087 was obtained from
\cite{2007yCat..90320844G}.}\label{figco5}
\end{figure}
\end{center}

We detect a bright single peaked line at 31~km~s$^{-1}$
(Figure~\ref{figco5}). HD85567 is located close to the galactic plane
in the direction of the Carina arm at $(l,b)=(282.6,-5.4)$. CO mappings
of the Carina region have not covered the position of HD85567
\citep{1999PASJ...51..751F,2001ApJ...553..274Z,2008PASJ...60.1297D} ,
therefore no comparison with known cloud velocities is available.

\subsubsection{HD~142527} 

HD~142527 is surrounded by a well-studied circumstellar disk that
harbors copious amounts of gas, as detected with the SMA
\citep{2008Ap&SS.313..101O,oberg2011}. The ASTE spectrum shows clear
contamination in the reference position (Figure~\ref{figco6}). The
emission coincides with the radial velocity of the star.\\

Our observations towards HD~142527 highlight the difficulties for using
CO(3-2) single dish observational techniques to detect unambiguously
circumstellar disc  emission in environments with high ambient
emission.  However, our follow up CO(6-5) observations using CHAMP+ in
APEX confirm warm CO emission in the position of the star
\citep{casassus13}, as well as the detection of a double-peaked
HCO+(4-3) line. \citep{degregoriomonsalvo}.  This may suggest that
warmer-gas or higher-density gas tracers may indeed be combined with
single dish observations to detect circumstellar gas emission from
protoplanetary disks. 

Recent ALMA imaging of this gas-rich disk has revealed the existence
of CO gas inside the dust cavity, as well as the presence of
gap-crossing HCO+ filaments \citep{2013Natur.493..191C}.

\subsubsection{HD~144668}

HD~144668 (V856 Sco, HR5999) is an A7~IVe star which exhibits a large
infrared excess ($L_{\rm IR}$/$L_{*}\sim 0.51$), with an age
of 2.8~Myr. Strong [O I] 63.2 $\mu$m emission towards HD~144668 is
detected \citep{meeus12}, indicative that its disk still has a large
reservoir of orbiting gas.\\

\begin{center}
\begin{figure}
\includegraphics[angle=0,scale=0.6]{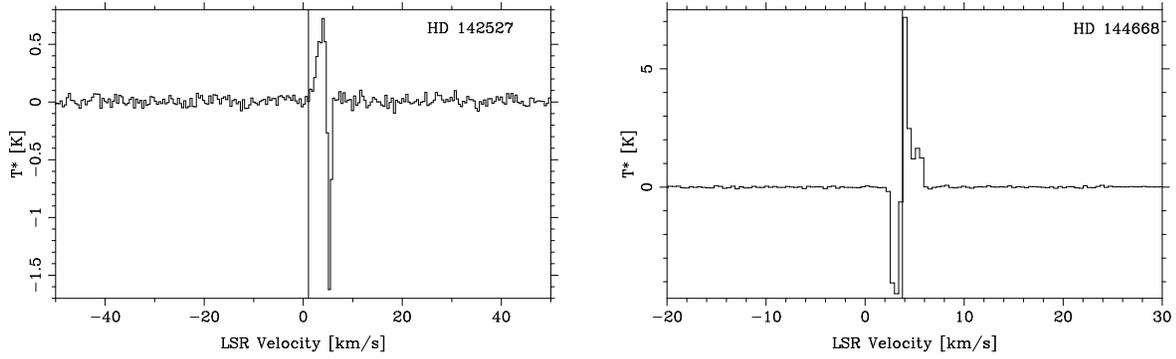}
\caption{Same as Figure~\ref{figco2}. The stellar
velocity for HD~142527 was obtained from
\cite{2007yCat..90320844G}.}\label{figco6}
\end{figure}
\end{center}

HD~144668 is located in the direction of the Lupus molecular clouds
\citep[Figure~\ref{figmilkyway};][]{1999PASJ...51..895H}.  The
emission in the CO(3-2) detection spectrum does coincide within errors
with the tabulated stellar velocity (Figure~\ref{figco6}). But also
coincides both in position and velocity with the C$^{18}$O core 27 in
\citet{1999PASJ...51..895H} ((l,b)=(339.567,933), $\rm{V}_{\rm{lsr}} =
4.12 ~$km s$^{-1}$) . This suggests that the CO emission observed
towards the star is dominated by diffuse cloud emission.  The
detection can be considered as an upper limit to the disk intrinsic CO
emission.

\subsubsection{HD~97048}

HD~97048 (CU Cha) is a well studied A0 star located in the direction
of the Chameleon I cloud
\citep[Figure~\ref{figmilkyway};][]{1998ApJ...507L..83M}. \citep{2009ApJ...695.1302M}
detected the Mid-IR H$_2$ S(1) line at 17~$\mu$m, but not the S(2) nor
the S(4) at 12 and 8~$\mu$m, respectively. \citet{2011A&A...533A..39C}
detect near-IR H$_2$ 1-0 S(1) using CRIRES on the VLT. Copious amounts
of CO J=19-17, [O I] 145.5 $\mu$m and [O I] 63.2 $\mu$m emission
towards HD~97048 are detected \citep{meeus12}. It is also one of the
three objects where [C II] 157.7 $\mu$m is detected.\\

The CO spectra is contaminated due to cloud emission in the reference
(Figure~\ref{figco7}), since the star lies on top of the Cha I
molecular cloud. The typical $^{12}$CO(1-0) cloud velocities in the
Cha I region are in the $\rm{V}_{\rm{lsr}} =2-6$ km s$^{-1}$ range
\citep{1998ApJ...507L..83M}, which concur with the (contaminated)
emission we detect.


\subsubsection{HD~132947}

\begin{center}
\begin{figure}
\includegraphics[angle=0,scale=0.6]{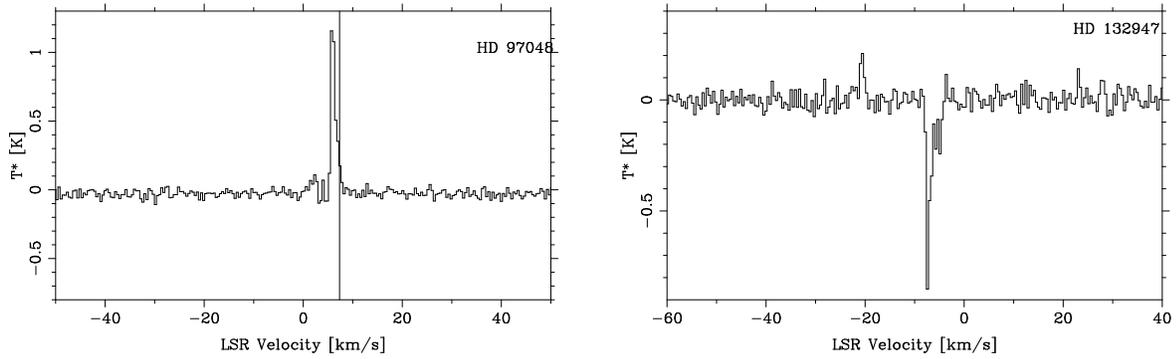}
\caption{Same as Figure~\ref{figco2}. The stellar velocity for HD~97048
  was obtained from \cite{2011A&A...533A..39C}. }\label{figco7}
\end{figure}
\end{center}

HD~132947 is located in the fourth galactic quadrant towards the
Circinus molecular cloud
\citep[Figure~\ref{figmilkyway};][]{2011ApJ...731...23S}. The NANTEN
$^{12}$CO maps show several velocity components ranging between
$\rm{V}_{\rm{lsr}} =-4$ and $-6.5$ km s$^{-1}$, in agreement with the
contamination seen in the reference position (which produces the false
absorption feature seen in the spectrum). A small emission feature is
seen near $\rm{V}_{\rm{lsr}} = -22 ~$km s$^{-1}$, but the lack of
information for the stellar velocity hampers any association to
possible circumstellar emission (Figure~\ref{figco7}).

\subsubsection{HD~250550 }

HD250550 is located towards the Gemini OB1 molecular clouds
\citep{1995ApJ...445..246C}. Based on existing $^{12}$CO(1-0) maps,
the velocity of the CO(3-2) emission detected in our spectra
(Figure~\ref{figco8}) coincides with the known cloud velocity in the
direction of HD250550 \citep[$\rm{V}_{\rm{lsr}} = 4.12 ~$km
  s$^{-1}$;][]{1995ApJ...445..246C}.

\subsubsection{HD~149914 }

\begin{center}
\begin{figure}
\includegraphics[angle=0,scale=0.6]{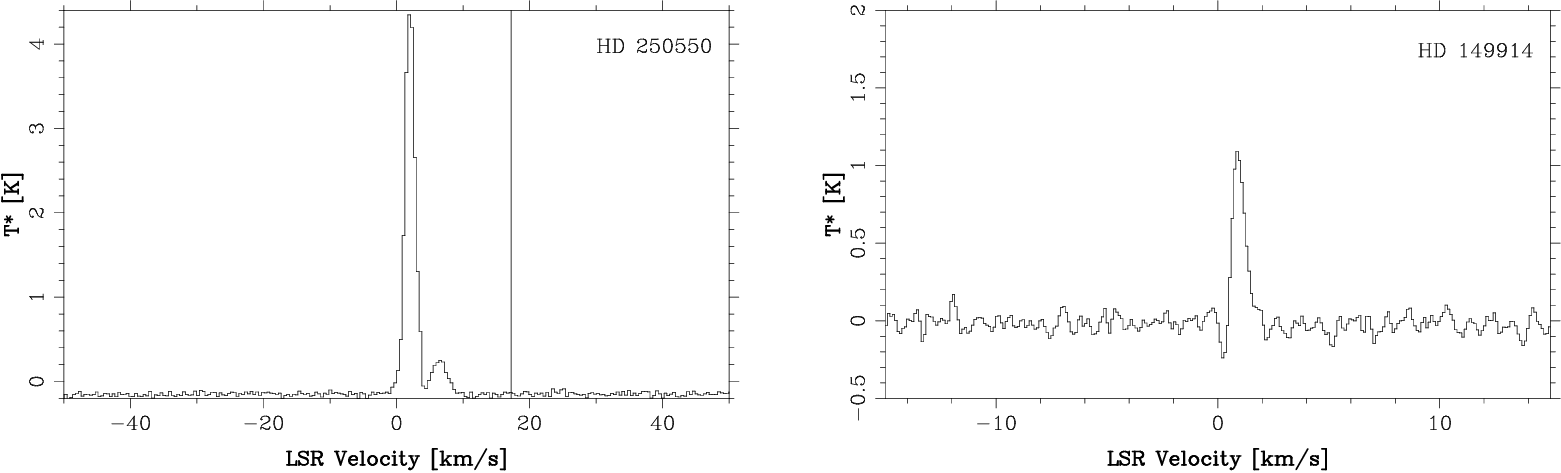}
\caption{Same as Figure~\ref{figco2}. The stellar
velocity for HD~250550 was obtained from
\cite{2007yCat..90320844G}.}\label{figco8}
\end{figure}
\end{center}

$^{12}$CO(3-2) emission is detected towards the B9.5 IV star HD~149914,
located towards the Ophiucus cloud \citep{dame01}. The spectrum shows
indications of contamination in the reference position near 0
km~s$^{-1}$ (Figure~\ref{figco8}).  At least four clouds nearby (1-3
degrees) with $\rm{V}_{\rm{lsr}} = 1-5 ~$km~s$^{-1}$ are
known \citep{2000ApJ...528..817T}.  The closest core to the
line-of-sight of HD~149914 is located at $(l,b)=(359.23, 18.63)$ 
has a main velocity component $\rm{V}_{\rm{lsr}} = 0.58 ~$km~s$^{-1}$
and width 0.92 km~s$^{-1}$.

\subsubsection{KK~Oph}

KK~Oph is a binary system located at the edge of the Pipe Nebula at
$(l,b)=(357.06,7.1)$, close to a C$^{18}$O core with
$\rm{V}_{\rm{lsr}} \sim 3.6 ~$km~s$^{-1}$ \citep{1999PASJ...51..871O}.
Cloud contamination is evident in our CO spectra
(Figure~\ref{figco9}), although the [O I] 63.2 $\mu$m detection
indicates that the KK~Oph system harbors a gas rich protoplanetary
disk \citep{meeus12}.

\subsubsection{VV~Ser}

\begin{center}
\begin{figure}
\includegraphics[angle=0,scale=0.6]{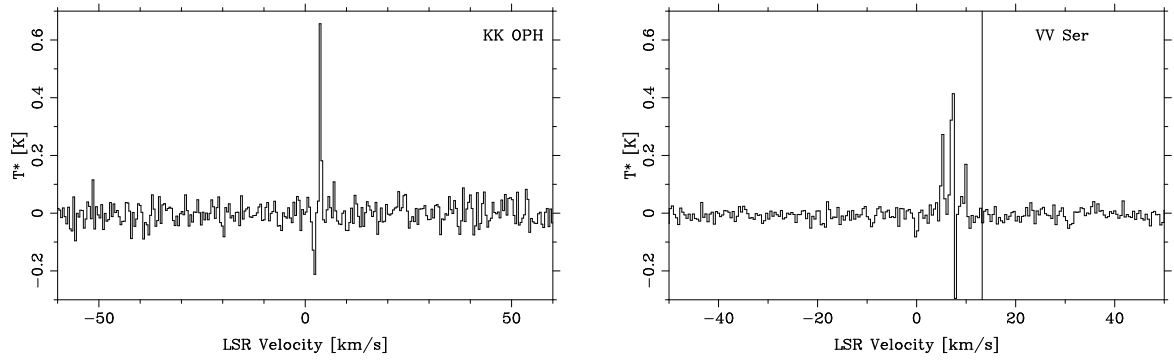}
\caption{Same as Figure~\ref{figco2}. The stellar
velocity for VV~Ser was obtained from \cite{acke05}}\label{figco9}
\end{figure}
\end{center}

Our observations show that VV~Ser is surrounded by extended CO(3-2)
that contaminates the reference position (Figure~\ref{figco9}), with
centroid velocity $\rm{V}_{\rm{lsr}} \sim 7~$km~s$^{-1}$.
\citet{alonso08} reported continuum detection (PdBI), corresponding to
a $4 \times 10^{-5}~M_\odot$ disk, and a CO(1-0) non-detection (at a
1~$\sigma$ noise limit of 0.3~K in 1~km~s$^{-1}$
channels). \citet{liu11} detect CO(2-1) and CO(3-2) using the KOSMA
3-meter telescope, and suggest that might come from three different
velocity components at 5.3, 7.3 and 9.5~km~s$^{-1}$, respectively. We
detect three $^{12}$CO(3-2) peaks at these same velocities. VV~Ser is
located in the direction of the Aquila Rift, which together with the
contaminated spectra indicates that the single-dish observations are
dominated by diffuse molecular cloud emission \citep{dame01}.

\begin{center}
\begin{figure}
\includegraphics[angle=0,scale=0.6]{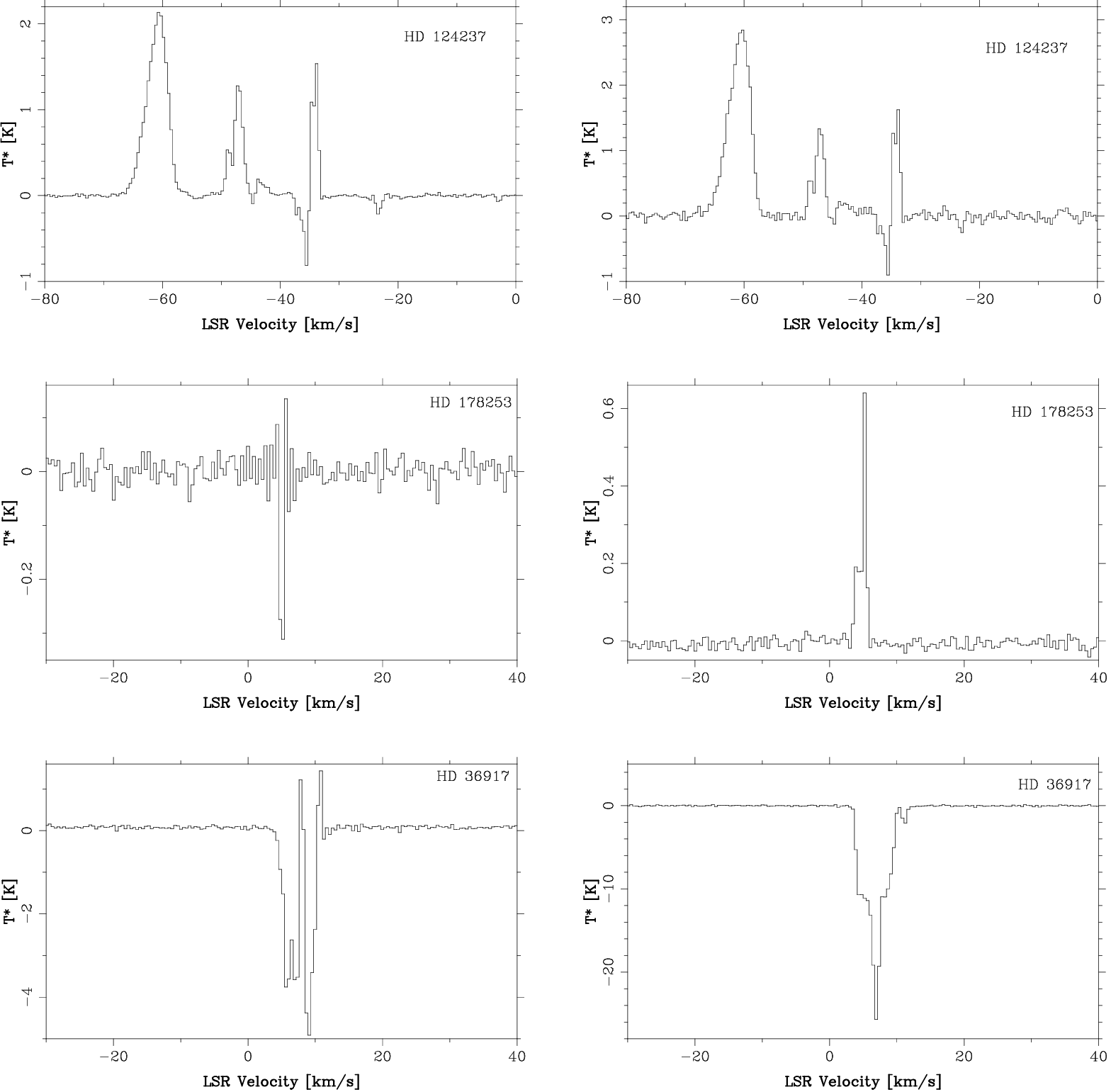}
\caption{Detections with background/foreground contamination in the
  reference position. The left-hand panels show the spectra obtained
  using a $-180$ arcsecond reference position, while the panels on the
  right present the spectra resulting from a $+180$ arcsecond
  reference position.  When available, the stellar velocity has been
  plotted with a vertical line (in $\rm{V}_{\rm{lsr}}$). Intensity
  units are in antenna temperature (T$^{*}_{A}$). }\label{figco2}
\end{figure}\label{figco2}
\end{center}

%
%

\begin{center}
\begin{figure}
\includegraphics[angle=0,scale=0.6]{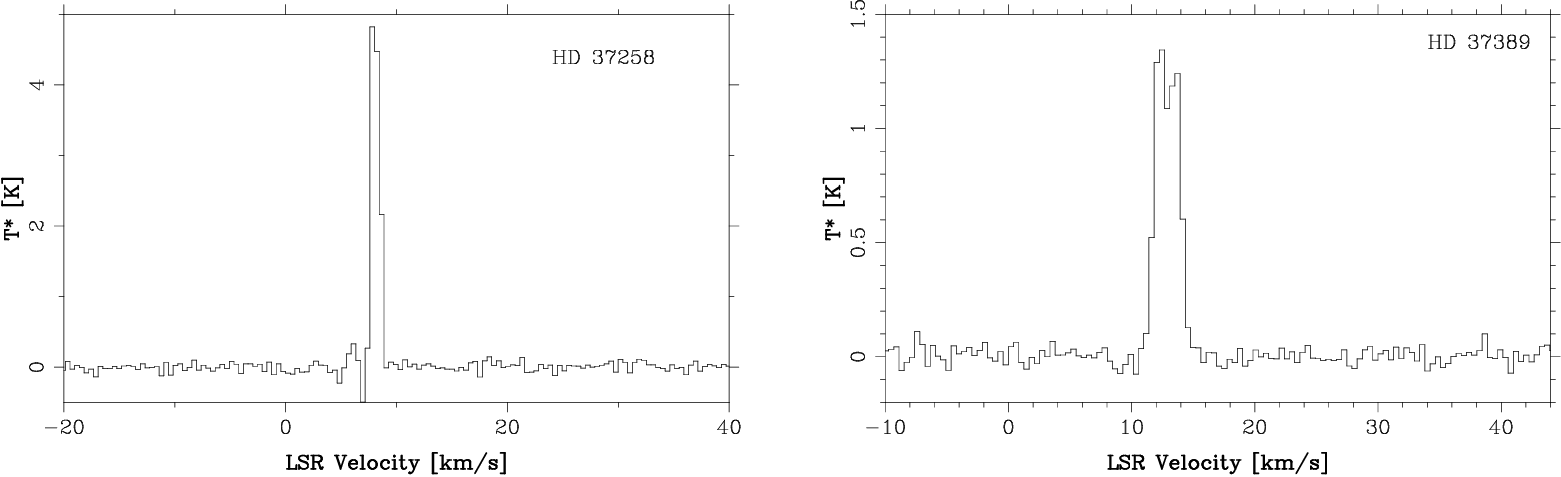}
\caption{Same as Figure~\ref{figco2}.}\label{figco4}
\end{figure}\label{figco4}
\end{center}

\clearpage

\subsection{CHAMP+ follow up observations}\label{achamp}

\begin{center}
\begin{figure}
\includegraphics[angle=0,scale=0.9]{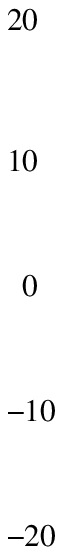}
\caption{CHAMP+ 7-horn footprint for HD~37389 in CO(6-5). The
  outermost axes show offset RA and DEC, in arcsec. The other axes
  give $T_A^\star$ in $y-$axis and $\rm{V}_\mathrm{lsr}$ is
  $x-$axis. The emission from the Orion Molecular Cloud is clearly
  traced by the CHAMP+ footprint.}\label{figchamp}
\end{figure}
\end{center}

Five sources with $^{12}$CO(3-2) detections (HD~142527, HD~178253,
HD~37389, RX J1842.9-3532 and VV Ser) were observed with the CHAMP+
dual color heterodyne array at $450\,\mu{\rm m}$ and $350\,\mu{\rm m}$
with $\sim 8''$ resolution. The 7-horn CHAMP+ receiver allows to map
the position of the star and its periphery simultaneously
\citep{2008SPIE.7020E..25G}, providing a 7-pixel map of the gas
emission around each star.  The CO(6-5)~691.4~GHz and C\,{\sc
  i}(2-1)~809~GHz were targeted in order to confirm the $^{12}$CO(3-2)
using the $^{12}$CO(6-5) line, and the C\,{\sc i}(2-1) to obtain
measures of the atomic gas content of the disks. These observations
are described in \citet{casassus13}. Only HD~142527 revealed a compact
$^{12}$CO(6-5) disk detection, and only C\,{\sc i}(2-1) upper
limits. HD~37389 shows clear contamination from extended emission,
particularly in the north-eastern horns of the CHAMP+ footprint
(Figure~\ref{figchamp}).

\begin{thebibliography}{}


\bibitem[Acke \& van den Ancker(2004)]{acke04} Acke, B., \& van den Ancker, M.~E.\ 2004, \aap, 426, 151 

\bibitem[Acke et al.(2005)]{acke05}{{Acke}, B. and {van den Ancker}, M.~E. and {Dullemond}, C.~P.},\ 2005, \aap, 436, 209 

\bibitem[Alexander(2013)]{2013arXiv1308.1791A} Alexander, R.\ 2013, arXiv:1308.1791 

\bibitem[Alonso-Albi et al.(2008)]{alonso08} Alonso-Albi, T., Fuente, A., Bachiller, R., et al.\ 2008, \apj, 680, 1289 

\bibitem[Andrews \& Williams(2005)]{2005ApJ...631.1134A} Andrews, S.~M., \& Williams, J.~P.\ 2005, \apj, 631, 1134 


\bibitem[Andrews et al.(2009)]{andrews2009} Andrews, S.~M., Wilner, D.~J., Hughes, A.~M., Qi, C., \& Dullemond, C.~P.\ 2009, \apj, 700, 1502 

\bibitem[Armitage(2011)]{armitage2011} Armitage, P.~J.\ 2011, \araa, 49, 195 




\bibitem[Balog et al.(2013)]{balog2013} Balog, Z., M{\"u}ller, T., Nielbock, M., et al.\ 2013, Experimental Astronomy, 38 

\bibitem[Barbier-Brossat \& Figon(2000)]{barbier00} Barbier-Brossat, M., \& Figon, P.\ 2000, \aaps, 142, 217 

\bibitem[Beichman et al.(1988)]{1988iras....1.....B} Beichman, C.~A.,  Neugebauer, G., Habing, H.~J., Clegg, P.~E., 
\& Chester, T.~J.\ 1988, Infrared astronomical satellite (IRAS) catalogs and atlases.~Volume 1: Explanatory supplement, 1

\bibitem[Birnstiel et al.(2010)]{Birnstiel2010} Birnstiel, T., Dullemond, C.~P., \& Brauer, F.\ 2010, \aap, 513, A79 


\bibitem[Blondel \& Djie(2006)]{2006A&A...456.1045B} Blondel, P.~F.~C., \& Djie, H.~R.~E.~T.~A.\ 2006, \aap, 456, 1045 



\bibitem[B{\"o}hm et al.(2004)]{bohm04} B{\"o}hm, T., Catala, C., Balona, L., \& Carter, B.\ 2004, \aap, 427, 907 

\bibitem[Boss(2001)]{boss01} Boss, A.~P.\ 2001, \apj, 563, 367 

\bibitem[Boss(2005)]{2005ApJ...629..535B} Boss, A.~P.\ 2005, \apj, 629, 535 

\bibitem[Boggess et al.(1978)]{1978Natur.275..372B} Boggess, A., Carr, F.~A., Evans, D.~C., et al.\ 1978, \nat, 275, 372 


\bibitem[Brittain et al.(2007)]{brittain07} Brittain, S.~D., Simon, T., Najita, J.~R., \& Rettig, T.~W.\ 2007, \apj, 659, 685 

\bibitem[Cabrit et al.(2006)]{2006A&A...452..897C} Cabrit, S., Pety, J., Pesenti, N., \& Dougados, C.\ 2006, \aap, 452, 897 


\bibitem[Carmona et al.(2008)]{carmona08} Carmona, A., van den Ancker, M.~E., Henning, T., et al.\ 2008, \aap, 477, 839 

\bibitem[Carmona(2010)]{2010EM&P..106...71C} Carmona, A.\ 2010, Earth Moon and Planets, 106, 71 

\bibitem[Carmona et al.(2011)]{2011A&A...533A..39C} Carmona, A., van der Plas, G., van den Ancker, M.~E., et al.\ 2011, \aap, 533, A39 

\bibitem[Carpenter et al.(1995)]{1995ApJ...445..246C} Carpenter, J.~M., Snell, R.~L., \& Schloerb, F.~P.\ 1995, \apj, 445, 246 

\bibitem[Casassus et al.(2013)]{casassus13} Casassus, S., Hales, A., de Gregorio, I., et al.\ 2013, \aap, 553, A64 

\bibitem[Casassus et al.(2013)]{2013Natur.493..191C} Casassus, S., van der Plas, G., M, S.~P., et al.\ 2013, \nat, 493, 191 


\bibitem[Collins et al.(2009)]{collins2009} Collins, K.~A., Grady, C.~A., Hamaguchi, K., et al.\ 2009, \apj, 697, 557 


\bibitem[Coulson et al.(1998)]{coulson98} Coulson, I.~M., Walther, D.~M., \& Dent, W.~R.~F.\ 1998, \mnras, 296, 934 

\bibitem[Cutri et al.(2003)]{2003yCat.2246....0C} Cutri, R.~M., Skrutskie, M.~F., van Dyk, S., et al.\ 2003, VizieR Online Data Catalog, 2246, 0 

\bibitem[Dahm \& Carpenter(2009)]{dahm09} Dahm, S.~E., \& Carpenter, J.~M.\ 2009, \aj, 137, 4024 

\bibitem[D'Alessio et al.(1999)]{dalessio99} D'Alessio, P., Calvet, N., Hartmann, L., Lizano, S., \& Cant{\'o}, J.\ 1999, \apj, 527, 893 

\bibitem[D'Alessio et al.(2006)]{2006ApJ...638..314D} D'Alessio, P., Calvet, N., Hartmann, L., Franco-Hern\'andez, R., \&Serv{\'{\i}}n, H.\ 2006, \apj, 638, 314 

\bibitem[Dame et al.(2001)]{dame01} Dame, T.~M., Hartmann, D., \& Thaddeus, P.\ 2001, \apj, 547, 792 

\bibitem[Dawson et al.(2008)]{2008PASJ...60.1297D} Dawson, J.~R., Kawamura, A., Mizuno, N., Onishi, T., \& Fukui, Y.\ 2008, \pasj, 60, 1297 

\bibitem[de Gregorio-Monsalvo et al.(2103)]{degregoriomonsalvo} De Gregorio-Monsalvo, I., et al., {{\it in preparation}}, \aap

\bibitem[Dent et al.(2005)]{dent05} Dent, W.~R.~F., Greaves, J.~S., \& Coulson, I.~M.\ 2005, \mnras, 359, 663 

\bibitem[de Winter et al.(2001)]{dewinter01} de Winter, D., van den Ancker, M.~E., Maira, A., et al.\ 2001, \aap, 380, 609 

\bibitem[de Zeeuw et al.(1999)]{dezeeuw99} de Zeeuw, P.~T., Hoogerwerf, R., de Bruijne, J.~H.~J., Brown, A.~G.~A., \& Blaauw, A.\ 1999, \aj, 117, 354 

\bibitem[Dobashi et al.(2005)]{dobashi05} Dobashi, K., Uehara, H., Kandori, R., et al.\ 2005, \pasj, 57, 1 

\bibitem[Draine \& Lee(1984)]{draine84} Draine, B.~T., \& Lee, H.~M.\ 1984, \apj, 285, 89 

\bibitem[Dunkin et al.(1997)]{dunkin97} Dunkin, S.~K., Barlow, 
M.~J., \& Ryan, S.~G.\ 1997, \mnras, 290, 165 

\bibitem[Dutrey et al.(1996)]{dutrey96} Dutrey, A., Guilloteau, S., Duvert, G., et al.\ 1996, \aap, 309, 493 

\bibitem[Fang et al.(2013)]{fang2013} Fang, M., van Boekel, R., Bouwman, J., et al.\ 2013, \aap, 549, A15 


\bibitem[Fajardo-Acosta et al.(1998a)]{fajardo98a} Fajardo-Acosta, S.~B., Telesco, C.~M., \& Knacke, R.~F.\ 1998a, \aj, 115, 2101 

\bibitem[Fajardo-Acosta et al.(1998b)]{fajardo98b} Fajardo-Acosta, S.~B., Stencel, R.~E., \& Backman, D.~E.\ 1998b, \apjl, 503, L193 


\bibitem[Fedele et al.(2013)]{fedele2013} Fedele, D., Bruderer, S., van Dishoeck, E.~F., et al.\ 2013, \aap, 559, A77 


\bibitem[Federman et al.(1980)]{Federman1980} Federman, S.~R., Glassgold, A.~E., Jenkins, E.~B., \& Shaya, E.~J.\ 1980, \apj, 242, 545 

\bibitem[Feigelson et al.(2003)]{feigelson03} Feigelson, E.~D., Lawson, W.~A., \& Garmire, G.~P.\ 2003, \apj, 599, 1207 


\bibitem[Folsom et al.(2012)]{} Folsom, C.P., Bagnulo, S., Wade, G.A., et al., 2012, MNRAS, 422, 2072 

\bibitem[Fukui et al.(1999)]{1999PASJ...51..751F} Fukui, Y., Onishi, T., Abe, R., et al.\ 1999, \pasj, 51, 751 

\bibitem[Fumel \& B{\"o}hm (2012)]{fumel2012} Fumel, A. \& B{\"o}hm, T.\ 2012, \aap, 540, A108 

\bibitem[Garcia et al.(2013)]{garcia2013} Garcia, P.~J.~V., Benisty, M., Dougados, C., et al.\ 2013, \mnras, 430, 1839 

\bibitem[ Garc\'{\i}a-L\'opez  et al.(2006)]{} Garc\'{\i}a-L\'opez, R., Natta, A., Testi, L., \& Habart, E.,
  2006, A\&A, 459, 837

\bibitem[Gontcharov(2007)]{2007yCat..90320844G} Gontcharov, G.~A.\ 2007, VizieR Online Data Catalog, 903, 20844

\bibitem[Guilloteau \& Lucas(2000)]{guilloteau00} Guilloteau, S., \& Lucas, R.\ 2000, Imaging at Radio through Submillimeter Wavelengths, 217, 299 

\bibitem[Guilloteau et al.(2013)]{guilloteau2013} Guilloteau, S., Di Folco, E., Dutrey, A., et al.\ 2013, \aap, 549, A92 



\bibitem[G{\"u}sten et al.(2006)]{gusten06} G{\"u}sten, R., Nyman, L.~{\AA}., Schilke, P., et al.\ 2006, \aap, 454, L13 


\bibitem[Grady et al.(2004)]{grady04} Grady, C.~A., Woodgate, B., Torres, C.~A.~O., et al.\ 2004, \apj, 608, 809 

\bibitem[Greaves et al.(2000)]{2000MNRAS.312L...1G} Greaves, J.~S., Coulson, I.~M., \& Holland, W.~S.\ 2000, \mnras, 312, L1 

\bibitem[G{\"u}sten et al.(2008)]{2008SPIE.7020E..25G} G{\"u}sten, R., Baryshev, A., Bell, A., et al.\ 2008, \procspie, 7020,  



\bibitem[Hamaguchi et al.(2005)]{hamaguchi05} Hamaguchi, K., Yamauchi, S., \& Koyama, K.\ 2005, \apj, 618, 360 

\bibitem[Hara et al.(1999)]{1999PASJ...51..895H} Hara, A., Tachihara, K., Mizuno, A., et al.\ 1999, \pasj, 51, 895 


\bibitem[Harju et al.(1993)]{harju93} Harju, J., Haikala, L.~K., Mattila, K., et al.\ 1993, \aap, 278, 569 

\bibitem[Hauck \& Mermilliod(1998)]{hauck1998} Hauck, B., \& Mermilliod, M.\ 1998, \aaps, 129, 431 


\bibitem[Henning et al.(1994)]{henning1994} Henning, T., Launhardt, R., Steinacker, J., \& Thamm, E.\ 1994, \aap, 291, 546 


\bibitem[Hillenbrand et al.(2008)]{hillenbrand08} Hillenbrand, L.~A., Carpenter, J.~M., Kim, J.~S., et al.\ 2008, \apj, 677, 630 

\bibitem[H{\o}g et al.(2000)]{hog2000} H{\o}g, E., Fabricius, C., Makarov, V.~V., et al.\ 2000, \aap, 355, L27 

\bibitem[Hollenbach et al.(2000)]{2000prpl.conf..401H} Hollenbach, D.~J., Yorke, H.~W., \& Johnstone, D.\ 2000, Protostars and Planets IV, 401 


\bibitem[Houk (1978)]{} Houk, N., 1978, Michigan catalogue of two-dimensional spectral types for the HD stars, Vizier on-line data III/51 



\bibitem[Hu et al.(1989)]{hu89} Hu, J.~Y., The, P.~S., \& de Winter, D.\ 1989, \aap, 208, 213 

\bibitem[Hughes et al.(2008a)]{2008ApJ...678.1119H} Hughes, A.~M., Wilner, D.~J., Qi, C., \& Hogerheijde, M.~R.\ 2008a, \apj, 678, 1119

\bibitem[Hughes et al.(2008b)]{hughes08} Hughes, A.~M., Wilner, D.~J., Kamp, I., \& Hogerheijde, M.~R.\ 2008b, \apj, 681, 626 

\bibitem[Hughes et al.(2010)]{hughes2010} Hughes, A.~M., Andrews, S.~M., Wilner, D.~J., et al.\ 2010, \aj, 140, 887

\bibitem[Ingleby et al.(2011)]{2011AJ....141..127I} Ingleby, L., Calvet, N., Hern{\'a}ndez, J., et al.\ 2011, \aj, 141, 127


\bibitem[Jonkheid et al.(2007)]{jonkheid07} Jonkheid, B., Dullemond, C.~P., Hogerheijde, M.~R., \& van Dishoeck, E.~F.\ 2007, \aap, 463, 203 

\bibitem[Juh{\'a}sz et al.(2010)]{juhasz10} Juh{\'a}sz, A., Bouwman, J., Henning, T., et al.\ 2010, \apj, 721, 431 

\bibitem[Kohno(2005)]{kohno05} Kohno, K.\ 2005, The Cool Universe: Observing Cosmic Dawn, 344, 242 

\bibitem[Kospal et al. (2013)]{kospal2013} Kospal, A. et al., {\it{submitted}}, 2013

\bibitem[Kraus et al.(2008)]{kraus08} Kraus, S., Hofmann, K.-H., Benisty, M., et al.\ 2008, \aap, 489, 1157 

\bibitem[Kun et al.(2000)]{kun00} Kun, M., Vink{\'o}, J., \& Szabados, L.\ 2000, \mnras, 319, 777 

\bibitem[Kurucz (1993)]{} Kurucz, R.L.,  1993, ATLAS9 Stellar Atmosphere Programs and 2 km/s grid. CD-ROM No. 13. Cambridge, Massachusetts,
  Smithsonian Astrophysical Observatory


\bibitem[Lee et al.(1996)]{lee1996} Lee, H.-H., Bettens, R.~P.~A., \& Herbst, E.\ 1996, \aaps, 119, 111 



\bibitem[Liu et al.(2011)]{liu11} Liu, T., Zhang, H., Wu, Y., Qin, S.-L., \& Miller, M.\ 2011, \apj, 734, 22 

\bibitem[Luna et al.(2008)]{luna08} Luna, R., Cox, N.~L.~J., Satorre, M.~A., et al.\ 2008, \aap, 480, 133


\bibitem[Lyo et al.(2008)]{lyo2008} Lyo, A.-R., Lawson, W.~A., 
\& Bessell, M.~S.\ 2008, \mnras, 389, 1461 

\bibitem[Malfait et  al.(1998)]{malfait98} Malfait, K., Bogaert, E., \& Waelkens, C.\ 1998, \aap, 331, 211 

\bibitem[{Mannings \& Barlow(1998)}]{mann98} Mannings, V., \& Barlow, M.~J. 1998, \apj, 497, 330


\bibitem[Martin-Za{\"i}di et al.(2008)]{zaidi08} Martin-Za{\"i}di, C., van Dishoeck, E.~F., Augereau, J.-C., Lagage, P.-O., \& Pantin, E.\ 2008, \aap, 489, 601 

\bibitem[Martin-Za{\"i}di et al.(2009)]{2009ApJ...695.1302M} Martin-Za{\"i}di, C., Habart, E., Augereau, J.-C., et al.\ 2009, \apj, 695, 
1302 

\bibitem[Meeus et al.(2001)]{meeus01} {{Meeus}, G., {Waters}, L.~B.~F.~M., {Bouwman}, J., 	{van den Ancker}, M.~E., {Waelkens}, C. and {Malfait}, K.}, 2001, \aap, 365, 476 


\bibitem[Meeus et al.(2012)]{meeus12} Meeus, G., Montesinos, B., Mendigut{\'{\i}}a, I., et al.\ 2012, \aap, 544, A78 




\bibitem[{{Meyer} {et~al.}(2007){Meyer}, {Backman}, {Weinberger}, \&
  {Wyatt}}]{meyer07}
{Meyer}, M.~R., {Backman}, D.~E., {Weinberger}, A.~J., \& {Wyatt}, M.~C. 2007,
  in Protostars and Planets V, ed. B.~{Reipurth}, D.~{Jewitt}, \& K.~{Keil},
  573

\bibitem[Mizuno et al.(1998)]{1998ApJ...507L..83M} Mizuno, A.,  Hayakawa, T., Yamaguchi, N., et al.\ 1998, \apjl, 507, L83

\bibitem[Mizuno et al.(2001)]{mizuno01} Mizuno, A., Yamaguchi, R., Tachihara, K., et al.\ 2001, \pasj, 53, 1071 

\bibitem[Moerchen et al.(2010)]{moerchen2010} Moerchen, M.~M., Telesco, C.~M., \& Packham, C.\ 2010, \apj, 723, 1418 

\bibitem[Montesinos et al.(2009)]{montesinos2009} Montesinos, B., Eiroa, C., Mora, A., \& Mer\'{\i}n, B., 2009, A\&A, 495, 901

\bibitem[Mo{\'o}r et al.(2011)]{moor2011} Mo{\'o}r, A., {\'A}brah{\'a}m, P., Juh{\'a}sz, A., et al.\ 2011, \apjl, 740, L7 

\bibitem[Nyman et al.(1992)]{nyman92} Nyman, L.-A., Booth, R.~S., Carlstrom, U., et al.\ 1992, \aaps, 93, 121

\bibitem[{\"O}berg et al.(2011)]{oberg2011} {\"O}berg, K.~I., Qi, C., Fogel, J.~K.~J., et al.\ 2011, \apj, 734, 98 

\bibitem[Ohashi(2008)]{2008Ap&SS.313..101O} Ohashi, N.\ 2008, \apss, 313, 101 

\bibitem[Onishi et al.(1999)]{1999PASJ...51..871O} Onishi, T., Kawamura, A., Abe, R., et al.\ 1999, \pasj, 51, 871 

\bibitem[Ortiz et al.(2005)]{ortiz05} Ortiz, R., Lorenz-Martins, S., Maciel, W.~J., \& Rangel, E.~M.\ 2005, \aap, 431, 565 


\bibitem[Owen et al.(2012)]{owen2012} Owen, J.~E., Clarke, C.~J., \& Ercolano, B.\ 2012, \mnras, 422, 1880 

\bibitem[Pani{\'c} \& Hogerheijde(2009)]{panic2009} Pani{\'c}, O., \& Hogerheijde, M.~R.\ 2009, \aap, 508, 707 

\bibitem[Pani{\'c} et al.(2010)]{2010A&A...519A.110P} Pani{\'c}, O., van Dishoeck, E.~F., Hogerheijde, M.~R., et al.\ 2010, \aap, 519, A110 

\bibitem[Pascucci et al.(2006)]{pascucci2006} Pascucci, I., Gorti, 
U., Hollenbach, D., et al.\ 2006, \apj, 651, 1177 

\bibitem[Pavlyuchenkov et al.(2007)]{Pavlyuchenkov2007} Pavlyuchenkov, Y., Semenov, D., Henning, T., et al.\ 2007, \apj, 669, 1262 

\bibitem[Pinte et al.(2006)]{pinte06} Pinte, C., M{\'e}nard, F., Duch{\^e}ne, G., \& Bastien, P.\ 2006, \aap, 459, 797 

\bibitem[Pollack et al.(1996)]{1996Icar..124...62P} Pollack, J.~B., Hubickyj, O., Bodenheimer, P., et al.\ 1996, Icarus, 124, 62 

\bibitem[Qi et al.(2011)]{qi2011} Qi, C., D'Alessio, P., {\"O}berg, K.~I., et al.\ 2011, \apj, 740, 84 



\bibitem[Redfield et al.(2007)]{redfield07} Redfield, S., Kessler-Silacci, J.~E., \& Cieza, L.~A.\ 2007, \apj, 661, 944 


\bibitem[Rhee et al.(2007a)]{rhee07a} Rhee, J.~H., Song, I., Zuckerman, B., \& McElwain, M.\ 2007a, \apj, 660, 1556 

\bibitem[Rhee et al.(2007b)]{rhee07b} Rhee, J.~H., Song, I.,\& Zuckerman, B.\ 2007b, \apj, 671, 616 

\bibitem[Rieke et al.(2005)]{rieke05} Rieke, G.~H., Su, K.~Y.~L., Stansberry, J.~A., et al.\ 2005, \apj, 620, 1010 

\bibitem[Sandell et al.(2011)]{sandell2011} Sandell, G., Weintraub, D.~A., \& Hamidouche, M.\ 2011, \apj, 727, 26 

\bibitem[Sch{\"u}tz et al.(2004)]{schutz04} Sch{\"u}tz, O., B{\"o}hnhardt, H., Pantin, E., et al.\ 2004, \aap, 424, 613 

\bibitem[Sch{\"u}tz et al.(2011)]{schutz2011} Sch{\"u}tz, O., Meeus, G., Carmona, A., Juh{\'a}sz, A., \& Sterzik, M.~F.\ 2011, \aap, 533, A54 


\bibitem[Shimoikura \& Dobashi(2011)]{2011ApJ...731...23S} Shimoikura, T., \& Dobashi, K.\ 2011, \apj, 731, 23 


\bibitem[Silverstone (2000)]{silv00} {Silverstone}, M.~D. 2000, Ph.D.~Thesis, University of California

\bibitem[Siringo et al.(2009)]{2009A&A...497..945S} Siringo, G., Kreysa, E., Kov{\'a}cs, A., et al.\ 2009, \aap, 497, 945 


\bibitem[Strai{\v z}ys et al.(1996)]{straizys96} Strai{\v z}ys, V., {\v C}ernis, K., \& Bartasiute, S.\ 1996, Baltic Astronomy, 5, 125


\bibitem[Sylvester \& Mannings(2000)]{sylvester00} Sylvester, R.~J.,  \& Mannings, V.\ 2000, \mnras, 313, 73

\bibitem[Tachihara et al.(2000)]{2000ApJ...528..817T} Tachihara, K., Mizuno, A., \& Fukui, Y.\ 2000, \apj, 528, 817 


\bibitem[Takeuchi \& Artymowicz(2001)]{takeuchi01} Takeuchi, T., \& Artymowicz, P.\ 2001, \apj, 557, 990 


\bibitem[Tatulli et al.(2007)]{tatulli07} Tatulli, E., Isella, A., Natta, A., et al.\ 2007, \aap, 464, 55 


\bibitem[Th\'e et al.(1994)]{the94} Th\'e, P.~S., de Winter, D., \& Perez, M.~R.\ 1994, \aaps, 104, 315 

\bibitem[Thi et al.(2001)]{thi01} Thi, W.~F., van Dishoeck, E.~F., Blake, G.~A., et al.\ 2001, \apj, 561, 1074 

\bibitem[van der Plas et al.(2008)]{2008A&A...485..487V} van der Plas, G., van den Ancker, M.~E., Fedele, D., et al.\ 2008, \aap, 485, 487 


\bibitem[van Kempen et al.(2007)]{vankempen07} van Kempen, T.~A., van Dishoeck, E.~F., Brinch, C., \& Hogerheijde, M.~R.\ 2007, \aap, 461, 983 

\bibitem[van Leeuwen(2007)]{venleeuwen2007} van Leeuwen, F.\ 2007, \aap, 474, 653 

\bibitem[Verhoeff et al.(2010)]{verhoeff2010} Verhoeff, A.~P., Min, M., Acke, B., et al.\ 2010, \aap, 516, A48 


\bibitem[Vieira et al.(2003)]{vieira03} Vieira, S.~L.~A., Corradi, W.~J.~B., Alencar, S.~H.~P., et al.\ 2003, \aj, 126, 2971 

\bibitem[Wade et al.(2007)]{wade07} Wade, G.~A., Bagnulo, S., Drouin, D., Landstreet, J.~D., \& Monin, D.\ 2007, \mnras, 376, 1145 

\bibitem[{{Walker} \& {Wolstencroft}(1988)}]{walker88}{Walker}, H.~J., \& {Wolstencroft}, R.~D. 1988, PASP, 100, 1509

\bibitem[Walsh et al.(2010)]{walsh2010} Walsh, C., Millar, T.~J., \& Nomura, H.\ 2010, \apj, 722, 1607

\bibitem[Weidenschilling(1977)]{weidenschilling77} Weidenschilling, S.~J.\ 1977, \mnras, 180, 57 

\bibitem[Williams \& Cieza(2011)]{williams11} Williams, J.~P., \& Cieza, L.~A.\ 2011, \araa, 49, 67 

\bibitem[Wyatt et al.(2005)]{wyatt05} Wyatt, M.~C., Greaves, J.~S., Dent, W.~R.~F., \& Coulson, I.~M.\ 2005, \apj, 620, 492 

\bibitem[Wyatt (2008)]{wyatt08} Wyatt, M.~C.\ 2008, \araa, 46, 339 

\bibitem[Wyatt et al.(2007)]{wyatt07} Wyatt, M.~C., Smith, R., Su, K.~Y.~L., et al.\ 2007, \apj, 663, 365 

\bibitem[Woitke et al.(2009)]{2009A&A...501..383W} Woitke, P., Kamp, I., \& Thi, W.-F.\ 2009, \aap, 501, 383 

\bibitem[Yamamura et al.(2011)]{yamamura2011} Yamamura, I., et al., 2011, AKARI/FIS All-Sky Survey Bright Source Catalogue Version 1.0 Release Note, http://www.ir.isas.jaxa.jp/AKARI/Observation/PSC/Public/

\bibitem[Yi et al. (2001)]{yi2001} Yi, S., Demarque, P., Kim, Y.-Ch., et al., 2001, ApJS, 136, 417

\bibitem[Zagorovsky et al.(2010)]{2010ApJ...720..923Z} Zagorovsky, K., Brandeker, A., \& Wu, Y.\ 2010, \apj, 720, 923

\bibitem[Zhang et al.(2001)]{2001ApJ...553..274Z} Zhang, X., Lee, Y., Bolatto, A., \& Stark, A.~A.\ 2001, \apj, 553, 274 

\bibitem[Zuckerman et al.(1995)]{zuckerman95} Zuckerman, B., Forveille, T., \& Kastner, J.~H.\ 1995, \nat, 373, 494 

\bibitem[Zuckerman (2001)]{zuckerman01} Zuckerman, B.\ 2001, \araa, 39, 549 

\end{thebibliography}
\end{document}